\documentclass[preprint,authoryear,12pt]{elsarticle}
\pdfoutput=1

\usepackage{graphics}
\usepackage{tikz}
\usepackage{ifpdf}
\ifpdf\usepackage{epstopdf}\fi
\graphicspath{{img/}}
\usepackage{url}

\usepackage{amssymb}
\usepackage{amsmath}

\usepackage{euscript}

\newcommand{\cF}{\EuScript{F}}

\newcommand{\cL}{\EuScript{L}}

\newcommand{\cP}{\EuScript{P}}

\newcommand{\rmT}{\mathrm{T}}

\newcommand{\tR}{\mathbb{R}}
\newcommand{\tS}{\mathbb{S}}

\newcommand{\tX}{\mathbb{X}}
\newcommand{\tY}{\mathbb{Y}}

\newcommand{\bfC}{\mathbf{C}}

\newcommand{\bfS}{\mathbf{S}}

\newcommand{\bfU}{\mathbf{U}}
\newcommand{\bfV}{\mathbf{V}}

\newcommand{\bfX}{\mathbf{X}}
\newcommand{\bfY}{\mathbf{Y}}
\newcommand{\bfZ}{\mathbf{Z}}

\newcommand{\bt}{\begin{theorem}}
\newcommand{\et}{\end{theorem}}
\newcommand{\bl}{\begin{lemma}}
\newcommand{\el}{\end{lemma}}
\newcommand{\bp}{\begin{proposition}}
\newcommand{\ep}{\end{proposition}}
\newcommand{\bc}{\begin{corollary}}
\newcommand{\ec}{\end{corollary}}

\newcommand{\bd}{\begin{definition}\rm}
\newcommand{\ed}{\end{definition}}
\newcommand{\bex}{\begin{example}\rm}
\newcommand{\eex}{\end{example}}
\newcommand{\br}{\begin{remark}\rm}
\newcommand{\er}{\end{remark}}

\newcommand{\btbh}{\begin{table}[!ht]}
\newcommand{\etb}{\end{table}}
\newcommand{\bfgh}{\begin{figure}[!ht]}
\newcommand{\efg}{\end{figure}}

\newcommand{\bea}{\begin{eqnarray*}}
\newcommand{\eea}{\end{eqnarray*}}
\newcommand{\be}{\begin{eqnarray}}
\newcommand{\ee}{\end{eqnarray}}

\newcommand{\suml}{\sum\limits}

\newcommand{\lm}{\lambda}
\def\wtilde{\widetilde}
\def\what{\widehat}

\def\spaceR{\mathsf{R}}
\newcommand{\bfw}{\mathbf{w}}

\def\last#1{{\underline{#1}}}
\def\first#1{{\mathstrut\overline{#1}}}

\def\sspan{\mathop{\mathrm{span}}}

\makeatletter
\def\adots{\mathinner{\mkern2mu\raise\p@\hbox{.}
\mkern2mu\raise4\p@\hbox{.}\mkern1mu
\raise7\p@\vbox{\kern7\p@\hbox{.}}\mkern1mu}}
\newcommand{\l@abcd}[2]{\hbox to\textwidth{#1\dotfill #2}}
\makeatother

\newtheorem{definition}{Definition}
\newtheorem{remark}{Remark}

\usepackage{float}
\makeatletter
\newcommand\floatc@codefragment[2]{{\@fs@cfont #1: #2}\par}
\newcommand\fs@codefragment{\def\@fs@cfont{\bfseries}\let\@fs@capt\floatc@codefragment
  \def\@fs@pre{\kern-8pt\relax}%
  \def\@fs@post{\kern-24pt\relax}%
  \def\@fs@mid{\kern8pt}%
  \let\@fs@iftopcapt\iftrue}
\makeatother
\floatstyle{codefragment}
\newfloat{fragment}{H}{lop}[section]
\floatname{fragment}{Fragment}

\usepackage{fancyvrb}
\DefineVerbatimEnvironment{Code}{Verbatim}{fontsize=\small}
\DefineVerbatimEnvironment{CodeInput}{Verbatim}{fontshape=sl}
\DefineVerbatimEnvironment{CodeOutput}{Verbatim}{}
\makeatletter
\newcommand\code{\bgroup\@makeother\_\@makeother\~\@makeother\$\@codex}
\def\@codex#1{{\normalfont\ttfamily\hyphenchar\font=-1 #1}\egroup}
\makeatother
\newcommand{\pkg}[1]{\textsc{#1}}
\def\R{{\normalfont\ttfamily R}}

\makeatletter
\def\verbatim@font{\small\ttfamily}
\makeatother

\journal{Computational Statistics \& Data Analysis}

\begin{document}

\begin{frontmatter}

\title{Basic Singular Spectrum Analysis and \\Forecasting with R}

\author[spbu]{Nina Golyandina}
\ead{nina@gistatgroup.com}

\author[spbu]{Anton Korobeynikov\corref{cor1}}
\ead{anton@korobeynikov.info}

\address[spbu]{Department of Statistical Modelling,
    Faculty of Mathematics and Mechanics, St.~Petersburg State University,
    Universitetsky pr 28, St.~Petersburg 198504, Russia}
\cortext[cor1]{corresponding author}

\begin{abstract}
    Singular Spectrum Analysis (SSA) is a powerful tool of analysis and
    forecasting of time series. In this paper we describe the main features of
    the \pkg{Rssa} package, which efficiently implements the SSA algorithms and
    methodology in \R. Analysis, forecasting and parameter estimation are
    demonstrated using case studies. These studies are supplemented with
    accompanying codes in \R.
\end{abstract}


\end{frontmatter}

\section{Introduction}
\label{sec:into}

Singular Spectrum Analysis (SSA) is a well-developed methodology of time series
analysis and forecasting which comprises many different but inter-linked
methods.  There are several books devoted to SSA
\citep{Elsner.Tsonis1996,Golyandina.etal2001,Golyandina.Zhigljavsky2012} as well
as many papers related to the theory of SSA and especially to various
applications of SSA (see \citet{Golyandina.Zhigljavsky2012} for references).
The scope of applications of SSA is very wide, from non-parametric time series
decomposition and filtration to parameter estimation and forecasting.

One of the differences between SSA and the methods of traditional time series
analysis is the fact that SSA and SSA-related methods can be applied to quite
different and not conventional for classical time series analysis problems such
as exploratory analysis for data-mining and parameter estimation in signal
processing, among others.  In this paper we mostly concentrate on exploratory
analysis by SSA; however, the tools for model construction and parameter
estimation are described too.  Despite no model is assumed before the SSA method
is applied, the so-called subspace-based model is constructed adaptively; the
corresponding class of time series includes time series governed by linear
recurrence relations in presence of noise.  Linear recurrence relations are
associated with autoregressive (AR) models. However, note that the AR model of
stationary processes is very different from the model of noisy time series
governed by LRR, which is associated with the model of deterministic signal
(generally, non-stationary) corrupted by noise. Therefore these models are
formally not comparable. For real-world time series whose models are unknown, AR
and SSA can be numerically compared, see e.g. \citet{Hassani.etal2009}.  The
essential difference between SSA and the majority of methods that analyze time
series with a trend and/or periodicities lies in the fact that SSA does not
require an a-priori model for trend as well as the a-priori knowledge of number
of periodicities and period values. Also, periodicities can be modulated by
different ways and therefore the type of model, additive or multiplicative, is not
necessary to be hold and taken into consideration.

Any method needs effective, comfortable and accessible implementation.  There
are many implementations of SSA. They differ by potential application areas,
implemented methods, interactive on non-interactive form, free or commercial
use, computer system (Windows, Unix, Mac), level of reliability and support.
The most known supported software packages implementing SSA are the following:
\begin{enumerate}
    \item
    \url{http://gistatgroup.com}:\\
    general-purpose interactive `Caterpillar'-SSA software (Windows) following
    the methodology described in \citet{Golyandina.etal2001,
        Golyandina.Zhigljavsky2012};

    \item
    \url{http://www.atmos.ucla.edu/tcd/ssa}:\\
    oriented mainly on climatic applications SSA-MTM Toolkit for spectral
    analysis \citep{Ghil.etal2002} (Unix) and its commercial extension kSpectra
    Toolkit (Mac), interactive;

    \item The commercial statistical software, SAS, includes SSA to its
    econometric extension SAS/ETS\circledR Software based on methodology of
    \citet{Golyandina.etal2001}.

    \item
    \url{http://cran.r-project.org/web/packages/Rssa}:\\
    \R{} package \pkg{Rssa} \citep{Korobeynikov2010}, a novel implementation of
    the main SSA procedures for major platforms, extensively developed.
\end{enumerate}

We consider the \pkg{Rssa} package as an efficient implementation of the main
SSA algorithms. This package also contains many visual tools which are useful
for making proper choice of SSA parameters and examination of the results.  At
present, \pkg{Rssa} is the only SSA implementation available from CRAN and
almost certainly the fastest implementation of the SSA.  Another important
feature of the package is its very close relation to the SSA methodology
thoroughly described in \citet{Golyandina.etal2001,Golyandina.Zhigljavsky2012}.
As a result of this, the use of the package is well theoretically and
methodologically supported. Note, however, that the package has been created
only recently (within the last 2 years) and therefore cannot be perfect. We are
aware about the ways of its further development and currently working on this
development.

The aim of this paper is to show how the methodology of the SSA analysis,
forecasting and parameter estimation can be implemented with the help of the
package \pkg{Rssa}.  Certainly, it is hard to study a method using only a short
description of it in a paper devoted to its \R{} implementation. Therefore we
refer the reader to \citet{Golyandina.etal2001,Golyandina.Zhigljavsky2012} which
contain detailed information on methodology and theory of SSA as well as
numerous references to applications of SSA to real-life time series and
comparison of SSA with other methods.

We start with a brief description of different aspects of the SSA methodology
(Section~\ref{sec:alg}) and present the structure of \pkg{Rssa} and features
of \pkg{Rssa} implementation in Section~\ref{sec:rssa}.  Sections~\ref{sec:alg}
and ~\ref{sec:rssa} provide some information necessary for the proper use of
\pkg{Rssa} functions and objects and proper application of \pkg{Rssa} for
analyzing real-life data.  The description of both the SSA methodology and the
\pkg{Rssa} package is not complete; much more information on SSA can be found in
\citet{Golyandina.etal2001, Golyandina.Zhigljavsky2012}, while technical
description of the \pkg{Rssa} functions can be found in the help files in the
package itself.

Sections~\ref{sec:decomp} and \ref{sec:for} contain examples of typical codes
for the analysis and forecasting, correspondingly. Each section contains a
simple example and also a case study. The examples demonstrate how to decompose
the time series into trend, periodic components and noise, how to choose SSA
parameters, how to estimate signal parameters (e.g. frequencies), how to perform
forecasting and check its accuracy.  In the sections with typical code fragments
we show how the functions from the \pkg{Rssa} package can be called and present
the codes for plotting figures; this is very important for making the right
choice of parameters and justification of the results.  However, we do not show
the figures themselves, since they require much space but can be easily obtained
by running the code.  Similar figures are shown in the sections devoted to case
studies.  The examples considered serve both for illustrating the use of
\pkg{Rssa} and for illustrating the theory and methodology discussed in
Section~\ref{sec:alg}. Therefore, we recommend to read Section~\ref{sec:alg}
together with running the typical codes and looking at the figures.

\section{SSA algorithms and methodology}
\label{sec:alg}
In this section we gather the information about SSA which is vital for
understanding the implementations of SSA and the ways SSA has to be used for the
analysis of real-life data.  One of basic tasks of SSA analysis is to decompose
the observed time series into the sum of interpretable components with no a
priori information about the time series structure.  Let us start with the
formal description of the algorithm.

\subsection{Algorithm of SSA analysis}
Consider a real-valued time series $\tX_N=(x_1,\ldots,x_{N})$ of length $N$.
Let $L$ ($1<L<N$) be some integer called {\em window length} and $K=N-L+1$.

The algorithm of SSA consists of two complementary stages: decomposition and
reconstruction.

\subsubsection{First Stage: Decomposition}
\label{subsect:decomp}
\paragraph{\bf 1st step: Embedding}
 To  perform the {\em embedding}
 we map  the original  time series into a sequence of
 lagged vectors of size $L$ by forming
$K=N-L+1$ {\em lagged vectors}
\bea
X_i=(x_{i},\ldots,x_{i+L-1})^\rmT,  \quad i=1\ldots,K.
\eea

The {\em trajectory matrix} of the series $\tX_N$ is
\be
\label{eq:traj_m}
\bfX=[X_1:\ldots:X_K]=(x_{ij})_{i,j=1}^{L,K}=\left(
\begin{array}{lllll}
x_1&x_2&x_3&\ldots&x_{K}\cr
x_2&x_3&x_4&\ldots&x_{K+1}\cr
x_3&x_4&x_5&\ldots&x_{K+2}\cr
\vdots&\vdots&\vdots&\ddots&\vdots\cr
x_{L}&x_{L+1}&x_{L+2}&\ldots&x_{N}\cr
\end{array}
\right).
\ee
There are two important properties of the trajectory matrix, namely, \\ (a) both
the rows and columns of $\bfX$ are subseries of the original series, and \\ (b)
$\bfX$ has equal elements on anti-diagonals and therefore the trajectory matrix
is Hankel.

\paragraph{\bf 2nd step: Decomposition}
Let $\{P_i\}_{i=1}^L$ be an orthonormal basis in $\spaceR^L$.
Consider the following decomposition of the trajectory matrix
\be
\label{eq:elem_matr}
\bfX=\sum_{i=1}^L P_i Q_i^\rmT=\bfX_1 + \ldots + \bfX_L,
\ee
where $Q_i=\bfX^\rmT P_i$, and define $\lambda_i=\|\bfX_i\|_\cF^2=\|Q_i\|^2$.

We consider two  choices of the basis $\{P_i\}_{i=1}^L$:
\begin{enumerate}
    \item[(A)] Basic: $\{P_i\}_{i=1}^L$ are eigenvectors of $\bfX \bfX^\rmT$;
    \item[(B)] Toeplitz: $\{P_i\}_{i=1}^L$ are eigenvectors of the matrix $\bfC$
    whose entries are
    $$
    c_{ij}=\frac{1}{N-|i-j|}
    \suml_{m=1}^{N-|i-j|}x_m x_{m+|i-j|}, \quad 1\leq i,j\leq L.$$
\end{enumerate}
In both cases the eigenvectors are ordered so that the corresponding eigenvalues
are placed in the decreasing order.

Let us remark that Case A corresponds to Singular Value Decomposition (SVD) of
$\bfX$, that is, $\bfX=\sum_i \sqrt{\lambda_i} U_i V_i^\rmT$, $P_i=U_i$ are left
singular vectors of $\bfX$, $Q_i=\sqrt{\lambda_i} V_i$, $V_i$ are called
\emph{factor vectors} or right singular vectors, $\lambda_i$ are eigenvalues of
$\bfX \bfX^\rmT$, therefore, $\lambda_1\geq\ldots\geq \lambda_L\geq 0$.

Note also that Case B is suitable only for the analysis of stationary time
series with zero mean (see e.g. \citet{Golyandina2010}). In the SSA literature
(A) is also called BK version, while (B) is called VG one.

The triple $(\sqrt{\lm_i},P_i,Q_i)$ will be called $i$th {\em
eigentriple} (abbreviated as ET).

\subsubsection{Second Stage: Reconstruction}
\index{Basic SSA stages!reconstruction}
\paragraph{\bf 3rd step: Eigentriple grouping}
Let $d=\max\{j:\ \lambda_j \neq 0\}$.
Once the expansion \eqref{eq:elem_matr} is obtained, the grouping procedure
partitions the set of indices $\{1,\ldots,d\}$ into $m$ disjoint subsets
$I_1,\ldots,I_m$.

Define $\bfX_I=\sum_{i\in I} \bfX_i$.
The expansion \eqref{eq:elem_matr} leads to the decomposition
\be
\label{eq:mexp_g}
\bfX=\bfX_{I_1}+\ldots+\bfX_{I_m}.
\ee
The procedure of choosing the sets $I_1,\ldots,I_m$ is called {\em eigentriple
    grouping}. If $m=d$ and $I_j=\{j\}$,
$j=1,\ldots,d$, then the corresponding grouping is called \emph{elementary}.
The choice of several leading eigentriples for Case A corresponds to the
approximation of the time series in view of the well-known optimality property
of the SVD.

\paragraph{\bf 4th step: Diagonal averaging}
At this step, we transform each matrix $\bfX_{I_j}$ of the grouped decomposition
(\ref{eq:mexp_g}) into a new series of length $N$.  Let $\bfY$ be an
$L\!\times\! K$ matrix with elements $y_{ij}$, $1\leq i\leq L$, $1\leq j\leq K$,
and let for simplicity $L\le K$.  By making the {\em diagonal averaging} we
transfer the matrix $\bfY$ into the series $(y_1,\ldots,y_N)$ using the formula
\bea
\label{eq:Hank}
  \widetilde{y}_{s} = \sum_{(l,k)\in A_s} y_{lk}\Big/| A_s|,
\eea
where $A_s=\{(l,k): l+k=s+1, 1\le l\le L, 1\le k\le K\}$ and
$|A_s|$ denotes the number of elements in the set $A_s$.
This corresponds to averaging  the matrix elements over the ``antidiagonals''.

Diagonal averaging \eqref{eq:Hank} applied to a resultant matrix $\bfX_{I_k}$
produces a \emph{reconstructed series} ${\wtilde \tX}^{(k)}=({\wtilde
    x}^{(k)}_1,\ldots,{\wtilde x}^{(k)}_N)$.  Therefore, the initial series
$(x_1,\ldots,x_N)$ is decomposed into a sum of $m$ reconstructed series:
\be
\label{eq:sexp_f}
  x_n = \suml_{k=1}^m{\wtilde x}^{(k)}_n,  \quad n=1,\ldots, N.
\ee
The reconstructed series produced by the elementary grouping will be called
\emph{elementary reconstructed series}.

\subsection{Separability and choice of parameters}
\label{sec:parameter_choice}
The very important question is how to choose parameters to construct the proper
decomposition of the observed time series and when this is possible.  Notion of
separability answer this question.  Separability of two time series
$\tX^{(1)}_N$ and $\tX^{(2)}_N$ signifies the possibility of extracting
$\tX^{(1)}_N$ from the observed sum $\tX^{(1)}_N + \tX^{(2)}_N$. SSA can
approximately separate, for example, signal and noise, sine waves with different
frequencies, trend and seasonality, and others \citep{Golyandina.etal2001,
    Golyandina.Zhigljavsky2012}.

If two time series are approximately separable, the problem of identification of
terms in \eqref{eq:elem_matr} corresponding to $\tX^{(1)}_N$ arises.  Time
series components can be identified on the base of the following principle: the
form of an eigenvector replicates the form of the time series component that
produces this eigenvector. Thus, graphs of eigenvectors can help in the process
of identification. Moreover, a sinusoid generates, exactly or approximately, two
sine wave components with the same frequency and the phase shift $\pi/2$.
Therefore, the scatterplot of a pair of eigenvectors, which produces a more or
less regular $T$-vertex polygon, can help to identify a sinusoid of period $T$.
For the problems of signal extraction, smoothing and noise reduction, several
leading eigentriples are chosen.

Very helpful information for separation is contained in the so-called
$\bfw$-correlation matrix. This is the matrix consisting of weighted
correlations between the reconstructed time series components. The weights
reflects the number of entries of the time series terms into its trajectory
matrix. Well separated components have small correlation whereas badly separated
components have large correlation. Therefore, looking at the $\bfw$-correlation
matrix one can find groups of correlated elementary reconstructed series and use
this information for the consequent grouping. One of the rules is not to include
into different groups the correlated components.

The conditions of (approximate) separability yield recommendations for the
choice of the window length $L$: it should be large enough ($L\sim N/2$) and if
we want to extract a periodic component with known period, then the window
lengths which are divisible by the period provide better separability. Choice of
parameters is discussed in \citet{Golyandina.etal2001} and
\citet{Golyandina2010}.  SSA with small $L$ performs smoothing of the series by
a filter of order $2L-1$ (see \citet{Golyandina.Zhigljavsky2012}), if we choose
a few leading eigentriples.  Generally, the choice of the window length is
important but the result is stable with respect to small changes of $L$.

If the time series has a complex structure, the so-called Sequential SSA is
recommended.  Sequential SSA consists of two stages, at the first stage the
trend is extracted with small window length and then periodic components are
detected and extracted from the residual with $L\sim N/2$.

If we use SSA as a model-free and exploratory technique, then the justification
of the decomposition is not formal; it is based on the separability theory and
interpretability of the results.  Real-time or batch processing by SSA is
possible if the class of series is specialized sufficiently allowing us to fix
the rule for choosing proper parameters.  For performing statistical testing, a
concrete model should be specified.

\subsection{Algorithms of SSA forecasting}

We formally describe the forecasting algorithms. For explanation see
\citet{Golyandina.etal2001}.

\subsubsection{Recurrent forecasting}
\label{sect:rec_for}

Let $I$ be the chosen set of eigentriples, $P_i\in \spaceR^L$, $i\in I$, be the
corresponding eigenvectors, $\last{P_i}$ be their first $L-1$ coordinates,
$\pi_i$ be the last coordinate of $P_i$, $\nu^2=\sum_i \pi_i^2$.  Define
$R=(a_{L-1},\ldots,a_1)^{\rmT}$ as \be
\label{eq:PVV_F}
R=\frac{1}{1-\nu^2}\suml_{i\in I}\pi_i\last{P_i}.
\ee

The \emph{recurrent forecasting algorithm} can be formulated as follows.
\begin{enumerate}
\item
The time series $\tY_{N+M}=(y_1,\ldots,y_{N+M})$ is defined by
\be
\label{eq:FORS}
y_i=\left\{
            \begin{array}{ll}
            \wtilde{x}_i &{\rm for\;}\; i=1,\ldots,N,\\
            \suml_{j=1}^{L-1} a_j y_{i-j} &{\rm for \;}\; i=N+1,\ldots,N+M.
            \end{array}
     \right.
\ee
\item
The numbers $y_{N+1},\ldots,y_{N+M}$ form the $M$ terms of the
recurrent forecast.
\end{enumerate}

Thus, the recurrent forecasting is performed by the direct use of the LRR with
coefficients $\{a_j, j=1,\ldots,L-1\}$.

\br
\label{rem:FOR_VEC}
Let us define the linear operator $\cP_{\mathrm{Rec}}:\spaceR^L\mapsto\spaceR^L$ by
the formula
\be
\label{eq:PA}
         \cP_{\mathrm{Rec}}Y=\left(\first{Y}\atop R^\rmT \first{Y}\right).
\ee
Set
\be
\label{eq:FOR_VEC}
Z_i=\left\{
            \begin{array}{ll}
            \wtilde{X}_i &{\rm for\;}\; i=1,\ldots,K,\\
            \cP_{\mathrm{Rec}} Z_{i-1} &{\rm for\;}\; i=K+1,\ldots,K+M.
            \end{array}
     \right.
\ee
It is easily seen that the matrix $\bfZ=[Z_1:\ldots:Z_{K+M}]$ is the trajectory
matrix of the series $\tY_{N+M}$. Therefore, (\ref{eq:FOR_VEC}) can
be regarded as the vector form of (\ref{eq:FORS}).
\er

\subsubsection{Vector forecasting}
\label{ssec:GEOM_F}
Denote ${\cL_r} =\sspan({P_i}, i\in I)$, $\what{X}_i$ the projection of the
lagged vector $X_i$ on $\cL_r$.  Consider the matrix
\be
\label{eq:fPI}
  \Pi =\last{\bfV} \last{\bfV}^\rmT + (1-\nu^2)R R^{\rm T},
\ee
where $\last{\bfV}=[\last{P_1}:\ldots:\last{P_r}]$ and $R$ is defined in
\eqref{eq:PVV_F}. The matrix $\Pi$ is the
matrix of the linear operator that performs the orthogonal projection
$\spaceR^{L-1}\mapsto \last{\cL_r}$, where $\last{\cL_r} =\sspan(\last{P_i},
i\in I)$.  Finally, we define the linear operator
$\cP_{\mathrm{Vec}}:\spaceR^L\mapsto \cL_r$ by the formula
\be
  \label{eq:PG}
  \cP_{\mathrm{Vec}}Y=\left(\Pi \first{Y}\atop  R^\rmT \first{Y}\right).
\ee

Let us formulate the \emph{vector forecasting algorithm}.
\begin{enumerate}
\item
In the notation above,  define  the vectors $Z_i$ as follows:
\be
\label{eq:V_FOR} Z_i=\left\{
            \begin{array}{ll}
                        \what{X}_i &{\rm for\;} \; i=1,\ldots,K,\\
                        \cP_{\mathrm{Vec}} Z_{i-1}&{\rm for\;} \; i=K+1,\ldots,K+M+L-1.\\
            \end{array}
     \right.
\ee
\item
By constructing the matrix $\bfZ=[Z_1:\ldots:Z_{K+M+L-1}]$ and making its
 diagonal
averaging we obtain the series $y_1,\ldots,y_{N+M+L-1}$.
\item
The numbers $y_{N+1},\ldots,y_{N+M}$ form the $M$ terms of the
vector forecast.
\end{enumerate}

In recurrent forecasting, we perform diagonal averaging to obtain the
reconstructed series and then apply the LRR. In vector forecasting these steps
are in a sense interchanged. The vector forecast is typically slightly more
stable but it has much larger computational cost than the recurrent forecast.

If the time series component is separated from the residual and is governed by
an LRR, both recurrent and vector forecasting coincide and provide the exact
continuation. In the case of approximate separability we obtain approximate
continuation.

Since LRRs provide the base for recurrent forecasting, let us consider time
series governed by LRRs in more details. It can be useful as from the viewpoint
of the parameter choice, as for understanding of the forecast behavior.

\subsection{Linear recurrence relations, time series of finite rank and roots}
\label{sec:lrr}
Let us consider the class of series that admit exact continuation by SSA
forecasting algorithms.  It is known that such series are governed by LRRs,
their trajectory matrices are rank-deficient, for these series the number of
non-zero terms in \eqref{eq:elem_matr} does not depend on $L$, and so on. This
class of series provides a natural model of the signal for SSA and especially
for forecasting. Let us introduce it formally.

\begin{definition}
A time series $\tS_N=\{s_i\}_{i=1}^{N}$ is {governed by a
linear recurrence relation (LRR)}, if there exist  $a_1,\ldots,a_t$ such that
\be
\label{eq:lrf}
s_{i+t}=\sum_{k=1}^t a_k s_{i+t-k},\ 1\leq i\leq N-t,\ a_t\neq 0, \ t<N.
\ee
The number $t$ is called the {order} of the LRR, $a_1,\ldots,a_t$ are the
coefficients of the LRR.  If $t=r$ is the minimal order of an LRR that governs
the time series $S_N$, then the corresponding LRR is called {minimal}.
\end{definition}

Time series is called \emph{time series of finite rank} $r$ if its
$L$-trajectory matrix has rank $r$ for any $L\ge r$ (recall that we always
assume $L\le K$).  Note that if the minimal LRR governing the signal $\tS_N$ has
order $r$ with $r<N/2$, then $\tS_N$ has rank $r$.

The minimal LRR is unique. Among all non-minimal LRRs of order $L-1$ the LRR
used in the recurrent SSA forecasting is the best (see
\citet{Golyandina.Zhigljavsky2012} for details).

\begin{definition}
    A polynomial $P_t(\mu)=\mu^t - \sum_{k=1}^t a_k \mu^{t-k}$ is called a
    {characteristic polynomial} of the LRR \eqref{eq:lrf}.
\end{definition}
Let the time series $\tS_{\infty}=(s_1,\ldots,s_n,\ldots)$ satisfy the LRR
\eqref{eq:lrf} with $a_t\neq 0$ and $i\geq 1$. Consider the characteristic
polynomial of the LRR \eqref{eq:lrf} and denote its different (complex) roots by
$\mu_1,\ldots,\mu_p$, where $p \leq t$.  All these roots are non-zero as
$a_t\neq 0$.  Let the multiplicity of the root $\mu_m$ be $k_m$, where $1\leq
m\leq p$ and $k_1+\ldots+k_p=t$.

It is well-known that the time series $\tS_{\infty}=(s_1,\ldots,s_n,\ldots)$
satisfies the LRR $\eqref{eq:lrf}$ for all $i\ge 0$ if and only if
\be
\label{eq:GEN_REQ}
s_n = \suml_{m=1}^p \left(\suml_{j=0}^{k_m-1} c_{mj} n^j\right) \mu_m^n,
\ee
where the complex coefficients $c_{mj}$ depend on the first $t$ points
$s_1,\ldots,s_{t}$.  For real-valued time series, \eqref{eq:GEN_REQ} implies
that the class of time series governed by the LRRs consists of sums of products
of polynomials, exponentials and sinusoids.

Rank of the series is equal to the number of non-zero terms in
\eqref{eq:GEN_REQ}.  For example, an exponentially-modulated sinusoid
$s_n=A e^{\alpha n}\sin(2\pi \omega n +\phi)$ is constructed from two
conjugate complex roots $\mu_{1,2}=e^{\alpha\pm \textsl{i} 2\pi \omega} = \rho
e^{\pm \textsl{i} 2\pi \omega}$ if its frequency
$\omega\in(0,0.5)$. Therefore, the rank of this exponentially-modulated sinusoid
is equal to 2.  The rank of the exponential is equal to 1, the rank of a linear
function corresponding to the root 1 of multiplicity 2 equals 2, and so on.

If we find the roots $\rho_j e^{\pm\textsl{i}2\pi\omega_j}$ of the characteristic
polynomial of the LRR governing the signal, then we can estimate the signal
parameters. For example, the frequency of an exponentially-modulated sinusoid
can be found using the argument of the corresponding conjugate roots, while root
modulus $\rho$ gives the exponential rate $\alpha=\ln \rho$.

If the LRR is not minimal, then only $r$ of the  roots correspond to the
signal. Other roots are extraneous and can influence the forecast. For example,
extraneous roots that have moduli larger than 1 can lead to instability.

\subsection{Estimation of frequencies}
Let $\tX_N=\tS_N+\tR_N$, where $s_n=\sum_{j=1}^r c_j \mu_j^n$ and the series
$\tS_N$ and $\tR_N$ are approximately separable for a given window length $L$.
As has been discussed above, the signal roots of the characteristic polynomial
of the forecasting LRR allow estimating the signal parameters $\mu_j$,
$j=1,\ldots,r$. However, we should somehow distinguish between signal and
extraneous roots.  Usually, the signal roots have maximal moduli (e.g. see
\citet{Usevich2010}). However, this is never guaranteed.  Therefore, the methods
that are able to separate the signal and extraneous roots could be very useful.

Let us describe one of these methods called ESPRIT \citep{Roy.Kailath1989}.
Denote by $I=\{i_1,\ldots,i_r\}$ the indices of the eigentriples which
correspond to $\tS_N$ (if $\tS_N$ is the signal then $I=\{1,\ldots,r\}$).  Set
$\bfU_r=[U_{i_1}:\ldots:U_{i_r}]$ and let $\last{\bfU_r}$ be the matrix with the
last row removed and $\last{\bfU_r}$ be the matrix with the first row
removed. Then $\mu_i$ can be estimated by the eigenvalues of the matrix
$\last{\bfU_r^\dag} \first{\bfU_r}$, where $\dag$ means pseudo-inversion.
Correspondingly, the estimated frequencies are the arguments of $\mu_i$.

There is an additional fast method of frequency estimation.  This method is
mostly used for the identification of the eigentriples at Grouping step.  Two
eigenvectors $U^{(1)}$ and $U^{(2)}$ produced by an exponentially-modulated sine
wave have similar form and their phases differ by $\pi/2$.  Let $A$ and $B$ be
defined by $a_n=\rho^n \sin(2\pi\omega n+\phi)$ and $b_n=\rho^n \cos(2\pi\omega
n+\phi)$.  Denote the angle between vectors by $\angle$. Then
$\omega=\angle\left(\left({a_1\atop b_1}\right), \left({a_2\atop
            b_2}\right)\right)/{2\pi}$.  Therefore, we can estimate the
frequency using the eigenvectors. Since the eigenvectors $U^{(1)}$ and $U^{(2)}$
do not have exactly the same form as $A$ and $B$, the sequence of angles
$\angle\left(\left({u^{(1)}_{i}\atop u^{(2)}_{i}}\right),
    \left({u^{(1)}_{i+1}\atop u^{(2)}_{i+1}}\right)\right)/{2\pi}$,
$i=1,\ldots,L-1$, can be considered and then the mean or median can be taken as
an estimate of the frequency (see \citet{Golyandina.etal2001} for details).  In
\pkg{Rssa}, the median is considered and the median of absolute deviations from
the median is taken as the measure of accuracy.

\subsection{Bootstrap confidence intervals}
Assume again $\tX_N=\tS_N+\tR_N$. Let us describe the construction of bootstrap
confidence intervals for the signal $\tS_N$ and its forecast assuming that the
signal has rank $r$ and the residuals are white %Gaussian
noise. The algorithm consists of the following steps.

\begin{itemize}
    \item Fix $L$, $I=\{1,\ldots,r\}$, apply SSA, reconstruct the signal and
    obtain the decomposition $\tX_N=\widetilde\tS_N+\widetilde\tR_N$.
    \item Fix $\widetilde\tS_N$, calculate the empirical distribution
    %calculate standard deviation $\sigma$
    of the residual $\widetilde\tR_N$.
    \item Simulate $Q$ independent copies $\widetilde\tR_{N,i}$, $i=1,\ldots,Q$,
    using the empirical distribution,
    %of white Gaussian noise with standard $\sigma$,
    construct
    $\widetilde\tX_{N,i}=\widetilde\tS_N+\widetilde\tR_{N,i}$.
    \item Apply SSA with the same $L$ and $I$ to $\widetilde\tX_{N,i}$,
    reconstruct the signal, then perform $M$-step ahead forecasting and
    obtain $\widetilde\tS_{N+M,i}$, $i=1,\ldots,Q$.
    \item For each time point $j$ consider the sample $\widetilde s_{j,i}$,
    $i=1,\ldots,Q$, and construct the bootstrap $\gamma$-confidence interval as
    the interval between $(1-\gamma)/2$- lower and upper sample quantiles.  The
    sample mean is called \emph{average bootstrap forecast}.
\end{itemize}

\subsection{Model and choice of parameters for forecasting}

While the SSA analysis generally does not require a model in advance, the SSA
forecasting does require a model. The model of the deterministic series that
admits the SSA forecasting is a signal, which is approximately governed by a
linear recurrence relation.  The SSA forecasting deals with sum of a signal and
a residual (maybe, noise), which should be approximately separated by SSA. This
is a rather general model, see Section~\ref{sec:parameter_choice} for examples
of approximately separable series and Section~\ref{sec:lrr} for description of
series governed by LRRs.  We should not specify the model precisely before
performing an SSA analysis; the dimension of the signal and the governing LRR
can be constructed by means of the SSA analysis itself. The associated
statistical testing of the constructed model can be performed by methods which
are not SSA-specific.

Basic rules for parameter choice in forecasting are generally the same as for
the reconstruction.  A considerable difference is that for forecasting a more
stable reconstruction can be even more important than the reconstruction
accuracy.  Also, simulations and theory \citep{Golyandina2010} show that it is
better to choose window length $L$ smaller than half of the time series length
$N$. One of the recommended values is $N/3$.

As a rule, recommendations are valid if the series approximately satisfies the
model of noisy time series governed by an LRR. Real-life time series always need
an additional analysis.

If the time series $\tX_N$ is long and has stable structure, the technique of
sliding forecasts can be applied.  We can choose the length $N_s$ of sliding
subseries, fix the window length $L$ and the group of indices $I$, choose the
forecasting horizon and then perform forecasts of subseries
$\tX_{i,i+N_s-1}=(x_i,\ldots,x_{i+N_s-1})$, $i=1,\ldots, N-N_s$.  The proper
choice of $L$ and $I$ corresponds to small average mean square error (MSE) of
forecasts. The choice of parameters allowing to obtain the minimal accuracy is
not necessary the best, since, for example, the stability with respect to small
changes of the window length may be more important.  For checking the stability
of forecasts the confidence intervals can be also useful.

If the time series is long but its structure can be changing in time, then the
estimation of forecast accuracy can help to understand how many of the most
recent points should be considered for forecasting.

\section{\pkg{Rssa} package}
\label{sec:rssa}
The main entry point to the package is \code{ssa} function which performs
Embedding step and (optionally, enabled by default) Decomposition step.
The function has the following signature:
\begin{verbatim}
ssa(x, L, ..., kind, svd.method, force.decompose = TRUE)
\end{verbatim}
Here \code{x} argument receives the input series, \code{L} specifies the window
length (equals to half of the series length by default), and \code{kind}
argument selects between different SSA algorithms.  In this paper we deal with
SSA for one-dimensional time series and therefore consider mostly the option
\code{kind="1d-ssa"} and shortly \code{kind="toeplitz-ssa"} (the option for
multivariate SSA \code{kind="2d-ssa"} is not considered).  Different
implementations of SVD can be selected via \code{svd.method} argument. These
implementations will be discussed later in Section~\ref{ssec:svd}. With default
value \code{force.decompose = TRUE} this function fulfills Decomposition stage
of the algorithm. All other arguments are passed to \code{decompose}
function. Usually this is \code{neig} argument which allows one to request the
desired number of eigentriples to compute (such request can be ignored depending
on the chosen SVD method).

The input time series for the \code{x} argument can be an ordinary numeric
vector (or a matrix for \code{kind="2d-ssa"}) or one of the standard time series
classes like \code{ts} or \code{zoo} from the package \pkg{zoo}
\citep{Zeileis.Grothendieck2005}. \pkg{Rssa} integrates well with various time
series classes which can be found on CRAN: the only requirement for the input
series is its convertibility to the standard numeric vector.  Note that the
\pkg{Rssa} package is used for the analysis and forecasting of equidistant
series. Thus, for example, the contents of the \code{index} attribute of a
\code{zoo} object is ignored on Decomposition stage.

The result of the \code{ssa} function is an SSA object which is the input for
the majority of other functions in the package. The contents of the object can
be viewed via \code{summary} function.

The function \code{reconstruct(x, groups)} is used to perform Reconstruction
stage.  The first argument is the SSA object. The second argument specifies the
eigentriple grouping \eqref{eq:mexp_g} and should be a list of vectors of
indices of the elementary series $I_{j}$. The return value of the function is
the list of reconstructed series corresponding to the input grouping. Note that
\code{reconstruct} function preserves all the attributes of the input
series. Thus, the reconstruction yields the \code{ts} object for \code{ts} input
series, etc. This behavior can be changed using the \code{drop} argument.

The principle of automatic calculation of necessary objects is used in the
implementation of the package. For example, if 10 eigentriples were calculated
while decomposing, then the user could still perform reconstruction by the first
15 components, since the decomposition will be automatically continued to
calculate 11--15 eigentriples. Also, the routines reuse the results of the
previous calculations as much as possible in order to save time (hence the
\code{cache} argument of many routines). For example, the elementary series once
calculated are stored inside the SSA object, so next time \code{reconstruct}
function might not need to calculate the resulting series from scratch. Also,
since SSA objects tend to occupy a decent amount of RAM, the functions and data
structures were designed to minimize the amount of memory copying.

Such efficient memory bookkeeping and invisible caching of the intermediate
results puts additional semantic burden on the SSA objects. In particular, SSA
objects effectively are \emph{references} and thus cannot be copied freely via
the standard assignment operator \code{<-}. Instead, the deep copy function
\code{clone} should be used. The internal cache can be freed via \code{cleanup}
routine.

The internals of an SSA object can be examined with the use of \code{\$}
operator. In particular, the following fields related to the
expansion~\eqref{eq:elem_matr} can be extracted out of SSA object: \code{lambda}
contains the eigenvalues ($\lambda_i$), \code{U} is a matrix with eigenvectors
($P_i$) in columns and \code{V} (might be \code{NULL}) is a matrix of factor
vectors ($Q_i/\|Q_i\|$).

\subsection{SVD methods}
\label{ssec:svd}
In many cases only a small number of leading eigentriples are of interest for
the SSA analysis. Thus the full SVD of the trajectory matrix can yield large
computational and memory space burden (here we consider the option
\code{type="1d-ssa"}). Instead, the so-called Truncated SVD can be used and only
a number of desired leading eigentriples can be computed. Four different SVD
implementations are available in \pkg{Rssa} and can be specified via
\code{svd.method} argument of \code{ssa} function:
\begin{itemize}
    \item \code{"nutrlan"}~--- Truncated SVD via thick restarted Lanczos
    bidiagonalization algorithm \citep{Yamazaki08}. The method internally
    calculates the eigenvalues and eigenvectors of the matrix $\bfX
    \bfX^\rmT$. Factor vectors are calculated on-fly during Reconstruction stage
    when necessary.
    \item \code{"propack"}~--- Implicitly restarted Lanczos bidiagonalization
    with partial reorthogonalization \citep{Larsen98}. The method calculates the
    truncated SVD of the trajectory matrix $\bfX$ (and thus calculates the
    factors vectors as well).
    \item \code{"eigen"} and \code{"svd"}~--- Full decomposition of the
    trajectory matrix using either eigendecomposition or SVD routines from
    \texttt{LAPACK} \citep{Lapack99}. These are basically the straightforward
    implementations of the basic SSA algorithm without any additional
    computational- and space complexity reductions via additional sophisticated
    algorithms. Note that both methods perform full decompositions and thus
    \code{neig} argument (which allows one to request desired number of
    eigentriples) is silently ignored for these methods.
\end{itemize}

Selecting the best method for performing SVD is difficult. However, there are
several easy rules of thumb which work very well in most situations.

First, it is unwise to use the Lanczos-based truncated SVD methods if the
trajectory matrix is small or ``wide''. This corresponds to small series lengths
(say, $N < 100$) or small window lengths ($L < 50$). Also, it is unwise to ask
for too many eigentriples: when more than half of window length eigentriples are
needed then it is better to use the full SVD instead of a truncated one.

SVD method \code{eigen} works best for small window lengths since in this case
the eigendecomposition of a small matrix needs to be computed.

Usually the \code{propack} method tends to be slightly faster and more
numerically stable than \code{nutrlan}, however, it may yield considerable
memory consumption when factor vectors are large. For example, for a time series
of length 87000 and window length 43500, the decomposition with \code{nutrlan}
method took 16 seconds while with \code{propack} it took only 13 seconds (we are
not aware about other besides \pkg{Rssa} implementations of the SSA algorithm
which can perform the decomposition with such window length at all). The memory
consumption for the latter method is as twice as the consumption of the
former. This difference is more important for multivariate version of SSA but
should not be a problem in our case.

By default \code{nutrlan} method is selected. However, \code{ssa} function
tries to correct the selection, when the chosen method is surely not the most
suitable.  In particular, for short series, small window length or large number
of desired eigentriples, the \code{eigen} method is automatically selected.

It should be note that truncated SVD implementations were extracted from
\pkg{Rssa} package into separate \pkg{svd} package and thus can be used
independently.

\subsection{Efficient implementation}
All the computation algorithms in the package are written with computation speed
in mind. The details of the algorithms can be found in
\citep{Korobeynikov2010}. Here we outline the computation complexities of
different SSA stages and the algorithms used.

\subsubsection{Basic SSA}
We should explicitly distinguish between specialized Lanczos-based SVD methods
(\code{nutrlan} and \code{propack}) and generic SVDs (\code{svd} and
\code{eigen}). The former can be made to exploit the special Hankel structure of
the trajectory matrix and thus reduce the computational and space complexity of
all the algorithms.

\begin{enumerate}
    \item Generic SVD methods:
    \begin{enumerate}
        \item Embedding step naturally has negligible computational
        complexity. Its space complexity is $O(LK)$. The worst case for generic
        SVD methods coincides with $L$ being equal to $N/2$ and the storage
        complexity is $O(N^{2})$.
        \item The Decomposition step computational complexity is $O(L^{3}
        +L^{2}K)$ for \code{eigen} method and $O(L^{2}K+LK^{2} + K^{3})$ for
        \code{svd} method \citep{Golub.VanLoan1996}. So, in the worst case $L
        \sim N/2$, the computation complexity is $O(N^{3})$ for both SVD methods.
        \item The computational complexity of Reconstruction stage depends on
        the upper bound for number of elementary series used. Let us denote this
        bound by $k$. Then the complexity of this stage is $O(kLK + k N)$ with
        the worst case being $O(kN^{2})$.
    \end{enumerate}
    \item Lanczos-based SVD methods:
    \begin{enumerate}
        \item Embedding step has $O(N \log{N})$ computational and $O(N)$
        storage complexity. The increased computational complexity is due to the
        additional preprocessing which is necessary for efficient algorithms
        used during Decomposition and Reconstruction stages. Note that no
        trajectory matrix is computed there, instead the representation via the
        so-called \emph{Toeplitz circulant} is used.
        \item The major speed-up can be seen during Decomposition step, since
        both truncated SVD and the special Hankel structure of the trajectory
        matrix contribute to the computation complexity. In particular, it can
        be shown \citep{Korobeynikov2010} that the multiplication of Hankel
        matrix by a vector can be viewed as a special case of convolution. The
        latter can be efficiently calculated by means of the Fast Fourier
        Transform (FFT).

        If $k$ eigentriples should be computed, then the complexity of such
        Hankel Truncated SVD is $O(kN\log{N} + k^{2}N)$ and does not depend on
        the window length.
        \item Reconstruction stage can be viewed as the formation of the elementary
        series and then taking  a sum of some of them depending on the grouping
        chosen. The computation of each elementary series, which is \emph{rank 1
            hankelization} can again be viewed as special form of
        convolution. Thus, the FFT-using Reconstruction is performed in $O(kN\log{N})$
        operations.
    \end{enumerate}
\end{enumerate}

All this explains the automatic choice of the SVD method described in the
previous section. From the comparison of the implementations we can conclude
that the Lanczos SVD-based methods work best when the window length $L$ is large
and the series length $N$ is not too small. Therefore, the Lanczos SVD-based
methods make it possible to achieve better separability by mean of the use of
large window length.

\subsubsection{Forecasting}
An efficient and stable implementation of the forecasting routines is necessary
not only for making the forecasts but also for studying the structure of the
series.

First, we should mention the procedure which calculates the roots of the
characteristic polynomial of the LRR. The task itself looks standard: we have to
calculate all the roots of the polynomial of the degree $L-1$, which can be
large, since is comparable with the series length.  Unfortunately, the standard
\R{} function \code{polyroot} which implements the classical Jenkins-Traub
algorithm \citep{Jenkins1970} often produces inaccurate results for the roots of
characteristic polynomials of LRRs. In \pkg{Rssa}, the roots are derived via
explicit eigenvalues calculation of the polynomial companion matrix.

Another computation-intensive routine is the vector forecast. The idea of the
method itself is simple: the resultant matrix of the reconstructed series should
be extended (by adding columns) while keeping the rank fixed. The classical
algorithm as in \citet{Golyandina.etal2001} involves the calculation of the
projections onto the space spanned by the selected eigenvectors.  For $p$-step
ahead forecasting, the complexities of doing such projections are
$O(k(p+L)L^{2})$, where $k$ is number of eigenvectors used for the
reconstruction. However, the problem of vector forecast can be reduced to the
ESPRIT-like system of linear equations. The effective solution of such system of
equations according to \citet{Badeau05} allows to reduce the complexity down to
$O(k^{2}(p+L))$.

\section{Basic SSA with R}
\label{sec:decomp}

\subsection{Typical code}
Let us consider the standard ``co2'' time series available in every \R{}
installation. The series depicts atmospheric concentrations of CO2 from Mauna
Loa Observatory, Hawaii, and contains 468 observations, monthly from 1959 to
1997 \citep{Keeling1997}. We choose this simple example for the code
demonstration only.

Code fragment~\ref{an:decomp} presents the typical code for construction of the
time series decomposition.

\begin{fragment}
\caption{``co2'': typical code of SSA analysis}
\begin{verbatim}
library(Rssa)
# Decomposition stage
s <- ssa(co2, L = 120)
# Reconstruction stage
# The results are the reconstructed series r$F1, r$F2, and r$F3
recon <- reconstruct(s, groups = list(c(1,4), c(2, 3), c(5, 6)))
# Calculate the residuals
res <- residuals(recon)
\end{verbatim}
\label{an:decomp}
\end{fragment}

The above code does not answer the question how to set \code{groups} to obtain
reasonable result. Proper grouping can be done looking on the diagnostic
plots. First, \code{plot} can be called on SSA object itself. Here \code{type}
argument can be used to select different plots available:
\begin{enumerate}
    \item \code{"values"} depicts eigenvalues (default);
    \item \code{"vectors"} shows 1D graphs of eigenvectors to detect trend
    components and saw-tooth component (if any);
    \item \code{"paired"} shows 2D graphs of eigenvectors to detect sine waves;
\end{enumerate}

Second, function \code{wcor} being applied to SSA object calculates the
$\bfw$-correlation matrix for the elementary reconstructed components. It can be
plotted in the standard way via \code{plot(wcor(s))}. Such picture can be used
to determine the separability points.

The use of these functions is summarized in the code of
Fragment~\ref{frg:co2dplots}. We omit the resultant figures; however,
the reader is recommended to run the code fragments and to look at the results
for understanding the methodology for this very simple example.
\begin{fragment}
    \caption{``co2'': diagnostic plots}
\begin{verbatim}
plot(s) # Eigenvalues
plot(s, type = "vectors") # Eigenvectors
plot(s, type = "paired") # Pairs of eigenvectors
plot(wcor(s)) # w-correlation matrix plot
\end{verbatim}
    \label{frg:co2dplots}
\end{fragment}

The result of the \code{reconstruct} function is at the same time a list with
components \code{F1}, \code{F2}, \dots, which contain the reconstructed series,
and the \emph{reconstruction object}, which can be conveniently plotted to see
the result of the reconstruction.

The \code{plot} method for the reconstruction object has two main arguments, which
configure the view of the resulting figure.
\begin{enumerate}
    \item \code{plot.method} argument might be \code{"matplot"} or
    \code{"native"} (default). In the former case all plotting is done via
    standard \code{matplot} function call. In the latter case the native
    plotting method of time series object is used for plotting (provides the
    best results for e.g. \code{ts} objects).
    \item \code{type} depicts whether the raw reconstructed series (argument
    value \code{"raw"}) or cumulative series \code{r\$F1}, \code{r\$F1+r\$F2}
    are to be plotted, and so on (argument value \code{"cumsum"}).
\end{enumerate}

So, in our case one can look at \code{plot(r)} for all reconstructed time series
separately, and \code{plot(r, type = "cumsum")} for cumulative series.

The \code{groups} argument of \code{wcor} function can be used to specify the
grouping used for reconstruction. The plot of such $\bfw$-correlation matrix can
be used to check the quality of separability. See Fragment~\ref{frg:co2rplots}
for example.

\begin{figure}[!htb]
    \centering
    \includegraphics{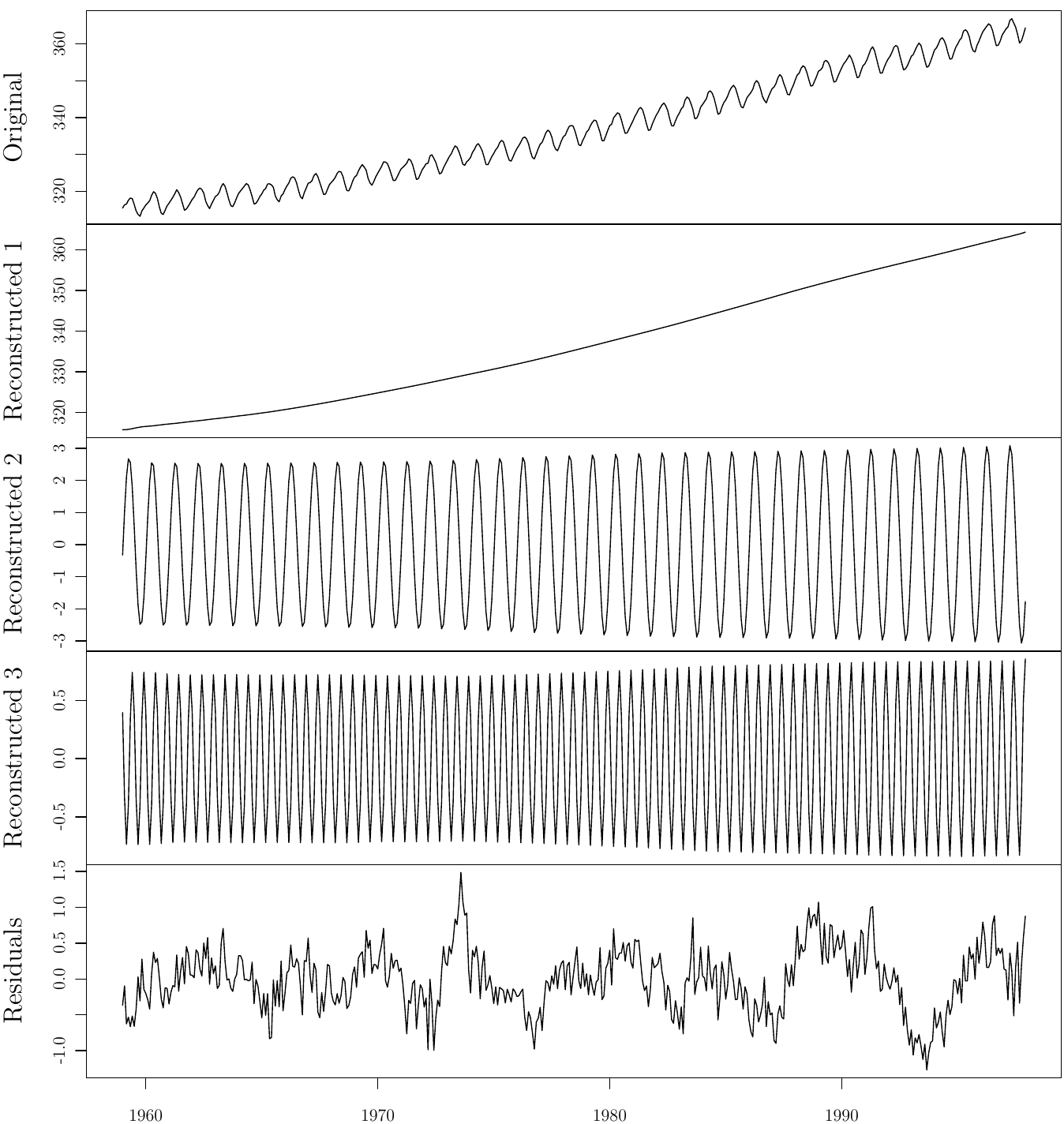}
    \caption{``co2'': full decomposition}
    \label{fig:co2_full}
\end{figure}

Additional logical arguments \code{add.residuals} and \code{add.original} can be
used to add the residuals and the original series to the reconstruction plots
(they are set to \code{TRUE} by default). In this way one can generate a figure
containing the decomposition into the sum of trend, seasonality and noise. The
result is depicted in Fig.~\ref{fig:co2_full}.

\begin{fragment}
\caption{``co2'': reconstruction plots}
\begin{verbatim}
# w-correlation matrix for reconstruction
plot(wcor(s, groups = list(c(1,4), c(2,3), c(5, 6))))
# Decomposition into trend + seasonality and noise
plot(recon)
\end{verbatim}
\label{frg:co2rplots}
\end{fragment}

Certainly, this resultant decomposition of the observed time series looks like a
trick, since we have not explained how the window length and the grouping have
been chosen. For ``co2'' series this can be done very easily and we address the
reader to the books \citep{Golyandina.etal2001, Golyandina.Zhigljavsky2012} for
detailed information and Section~\ref{sec:parameter_choice} for short
description of the principles of parameter choice. In fact, ``co2'' series does
contain two additional sine-wave components, which are hidden inside the
residuals. We leave the procedure of finding these components as an exercise for
the reader.

Below we consider a more complicated example with explanation of parameter
choice, using the two-stage Sequential SSA.

\subsection{Case study}

Let us analyze the time series ``MotorVehicle'' which contains monthly data of
total domestic and foreign car sales in the USA \citep{MotorVehicle2012}, from
1967 to 2012, January.

We start with the code resulting in the time series decomposition, then show the
graphs and comment on the logic of the investigation.

We will assume that \pkg{Rssa} package is already loaded. The series is
available from the package and can be loaded via \code{data(MotorVehicle)}
command. Total series length is 541.

\begin{figure}[!htb]
    \centering
    \includegraphics{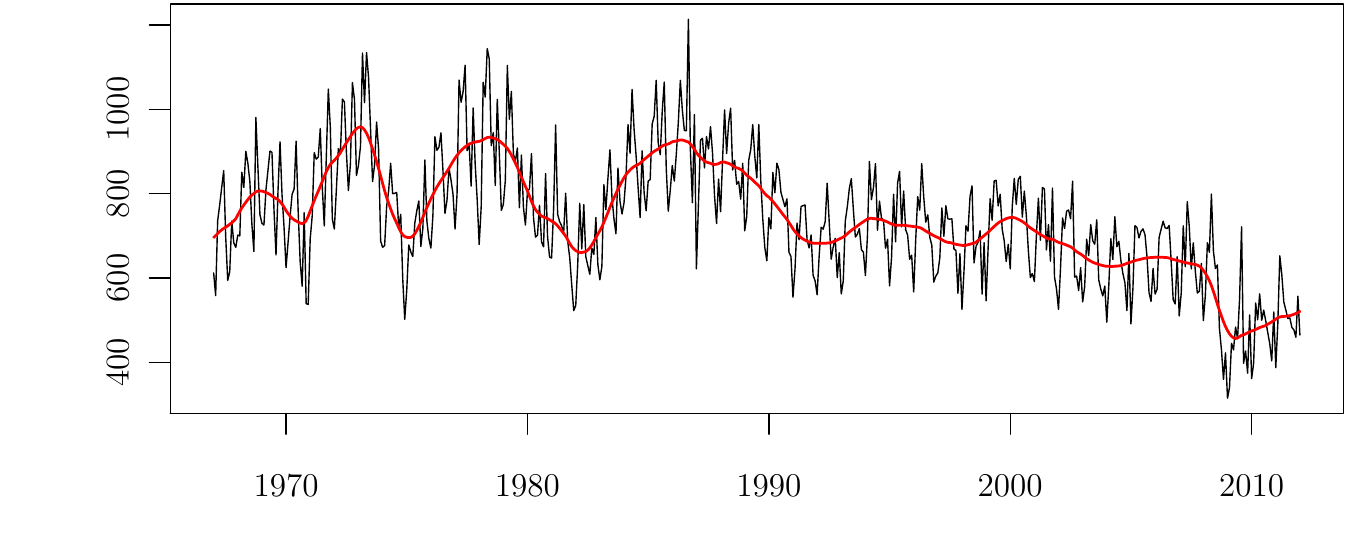}
    \caption{``MotorVehicle'', 1st stage: initial series and estimated trend ($L=12$, ET1)}
    \label{fig:sertrend}
\end{figure}

Fig.~\ref{fig:sertrend} shows that the form of trend is complex. This causes
impossibility to obtain the full decomposition of the time series at
once. Therefore, let us perform the decomposition sequentially.  First, let us
extract trend. Since for such changing form of the trend its extraction is
similar to smoothing, we start with choosing minimally possible window length
which in this case is $L=12$.  The reason for this choice of window length is
similar to that in moving averaging procedure: for smoothing the time series
containing a periodic component, the window length should be divisible by the
period.

Fragment~\ref{an:1dec} performs the decomposition and displays the information
about the resulting ssa-object.

\begin{fragment}
\caption{``MotorVehicle'', 1st stage: decomposition}
\begin{verbatim}
s1 <- ssa(MotorVehicle, L=12)
# Look inside 's' object to see, what is available.
summary(s1)
\end{verbatim}
\label{an:1dec}
\end{fragment}

This is an example of the output of \code{summary(s1)}:
\begin{verbatim}
Call:
ssa(x = MotorVehicle, L = 12)
Series length: 541,    Window length: 12,    SVD method: eigen
Computed:
Eigenvalues: 12,       Eigenvectors: 12,     Factor vectors: 0
Pre-cached: 0 elementary series (0 MiB)
Overall memory consumption (estimate): 0.005791 MiB
\end{verbatim}

The SVD method ``eigen'' was chosen by default, since the window length is small
and therefore fast SVD methods are not effective.  Since the pre-caching is
implemented in \pkg{Rssa}, it is important to know what the elements are already
calculated. You can see that there are 12 eigenvectors and 0 elementary
reconstruction components.

Now let us look at the decomposition results in Fragment~\ref{an:1decvis} for the
component identification.

\begin{fragment}
\caption{``MotorVehicle'', 1st stage: visual information for grouping}
\begin{verbatim}
# Plot of eigenvalues
plot(s1)
# Plot of eigenvectors
# Here 'idx' argument denotes the indices of vectors of interest
plot(s1, type = "vectors", idx=1:6)
# Plot of elementary reconstructed series
# Here 'groups' argument specifies the grouping
plot(s1, type = "series", groups = as.list(1:6))
\end{verbatim}
\label{an:1decvis}
\end{fragment}

Note that the plot of eigenvalues does not need additional calculations due to
pre-caching, while the plot of elementary reconstructed components needs
additional time for calculations (though such calculations are performed only
once for a given set of elementary components). The repeated call of
\code{summary(s1)} shows that \code{Precached: 6 elementary series (0.02497
    MiB)}.

\begin{figure}[htb]
    \centering
    \includegraphics{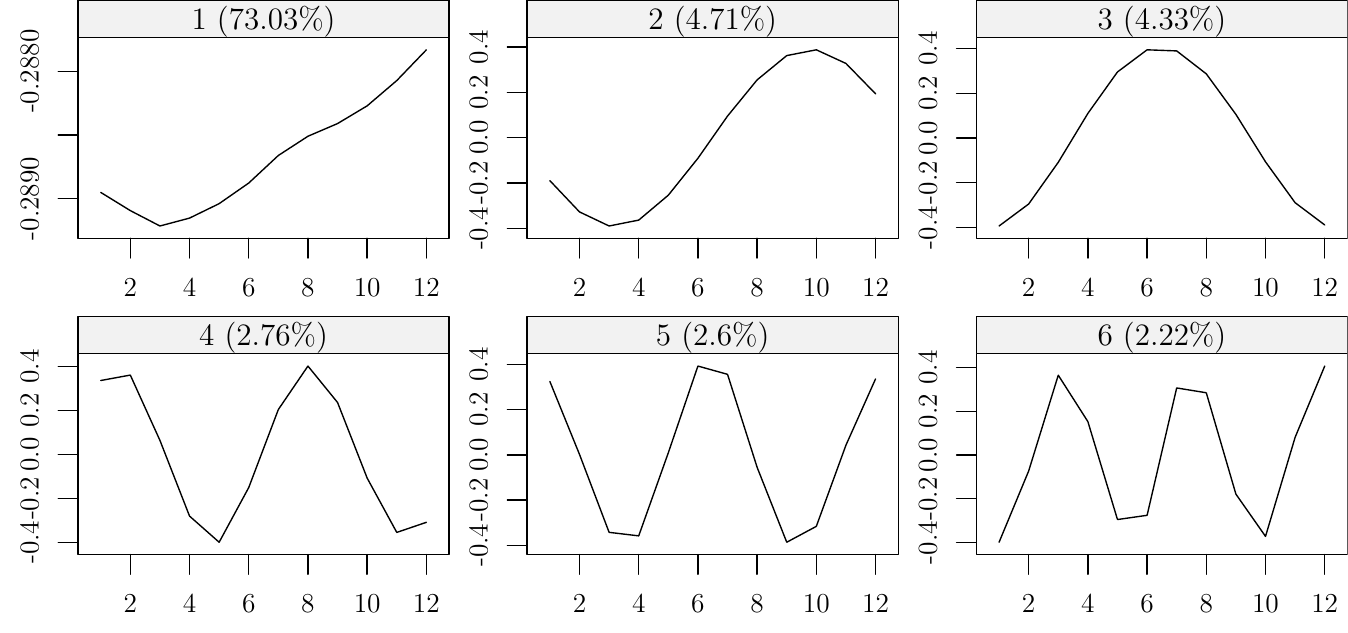}
    \caption{``MotorVehicle'', 1st stage: eigenvectors ($L=12$)}
    \label{fig:eigvec}
\end{figure}

\begin{figure}[htb]
    \centering
    \includegraphics{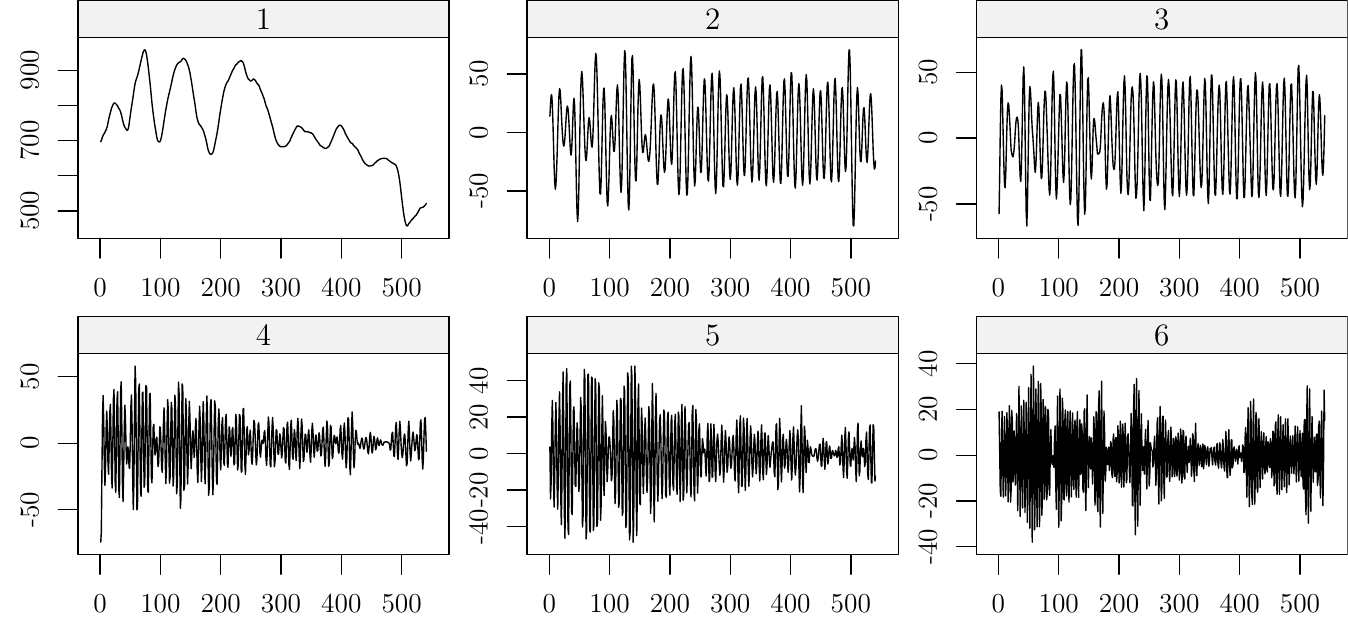}
    \caption{``MotorVehicle'', 1st stage:  elementary reconstructed series ($L=12$)}
    \label{fig:elseries}
\end{figure}

The graph of eigenvalues is not informative here and just reflects a large
contribution of the leading eigentriple. Fig.~\ref{fig:eigvec} shows the form of
the six leading eigenvectors. The leading eigenvector has almost constant
coordinates and therefore it corresponds to a pure smoothing by the Bartlett
filter (see \citet{Golyandina.etal2012} and \citet{Golyandina.Zhigljavsky2012}).
The result of reconstruction by each of the six eigentriples is depicted in
Fig.~\ref{fig:elseries}. Both figures confirm that the first eigentriple
corresponds to the trend, the other eigentriples contain high-frequency
components and therefore are not related to the trend.  The trend from
Fig.~\ref{fig:sertrend} is exactly the trend produced by one leading eigentriple
and coincides with the first reconstructed component in
Fig.~\ref{fig:elseries}. The trend can be reconstructed by the code from
Fragment~\ref{an:1rec}.

\begin{fragment}
\caption{``MotorVehicle'', 1st stage: reconstruction}
\begin{verbatim}
res1 <- reconstruct(s, groups = list(1))
trend <- res1$F1
\end{verbatim}
\label{an:1rec}
\end{fragment}

\begin{figure}[htb]
    \centering
    \includegraphics{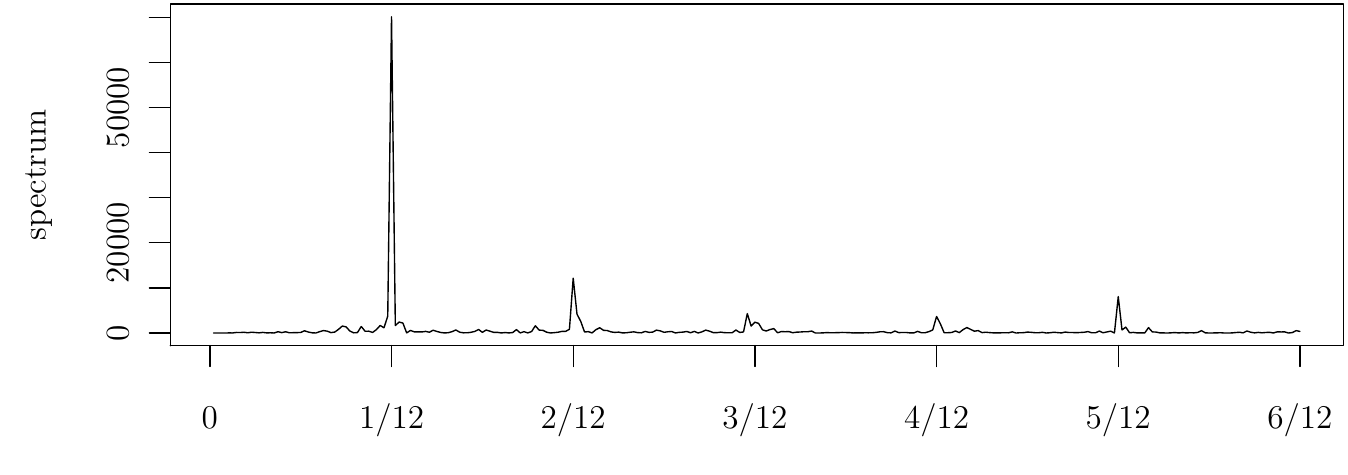}
    \caption{``MotorVehicle'', 2nd stage: periodogram of the series (i.e., of the residual at the 1st stage)}
    \label{fig:restrpgram}
\end{figure}

We have now extracted the trend and therefore the next stage is the extraction
of seasonality from the residual obtained by the command \code{res.trend <-
    residuals(res1)}.  First, let us look at the periodogram
(Fig.~\ref{fig:restrpgram}) by call \code{spec.pgram(res.trend, detrend = FALSE,
    log = "no")}.  We see that the seasonality consists of sine waves with
periods 12, 6, 4, 3, 2.4.  Let us extract them by the SSA.

For better separability, we take the window length $L=264$ as the maximal window
length $L$ such that $L \leq N/2$ and $L$ is divisible by 12.

\begin{fragment}
\caption{``MotorVehicle'', 2nd stage: decomposition and visual information}
\begin{verbatim}
s2 <- ssa(res.trend, L=264)
plot(s2)
plot(s2, type = "paired", idx = 1:12, plot.contrib = FALSE)
# Calculate the w-correlation matrix using first 30 components.
# Here groups argument as usual denotes the grouping used.
w <- wcor(s, groups = as.list(1:30))
plot(w)
\end{verbatim}
\label{an:2devis}
\end{fragment}
The code \code{summary(s2)} shows the chosen method \code{SVD method:\ nutrlan}
and the number of calculated eigenvalues and eigenvectors, which is 50
(default).

\begin{figure}[htb]
    \centering
    \includegraphics{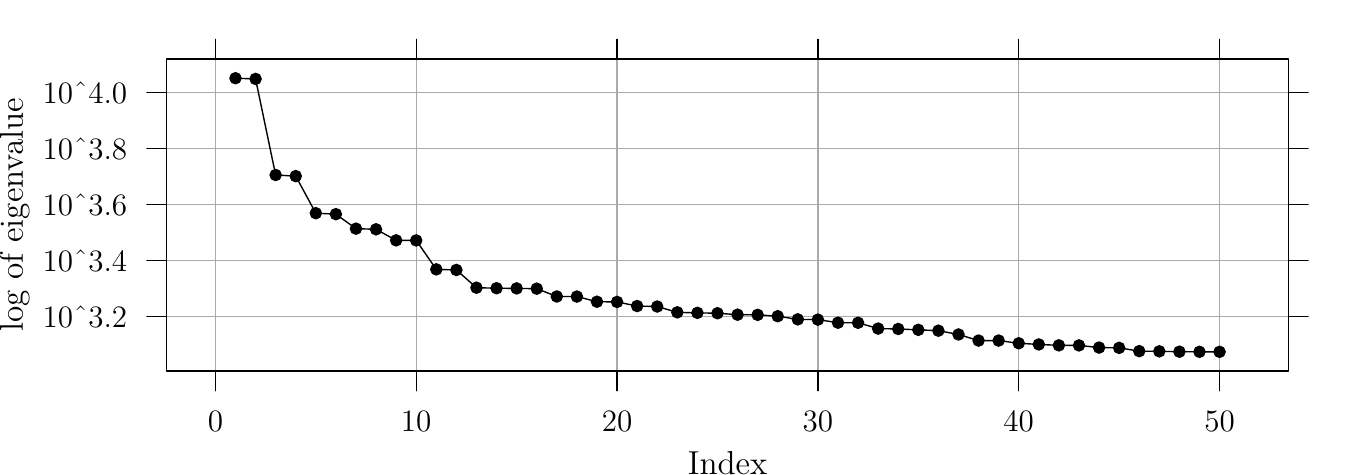}
    \caption{``MotorVehicle'', 2nd stage: eigenvalues ($L=264$)}
    \label{fig:eigvseason}
\end{figure}

\begin{figure}[!htb]
    \centering
    \includegraphics{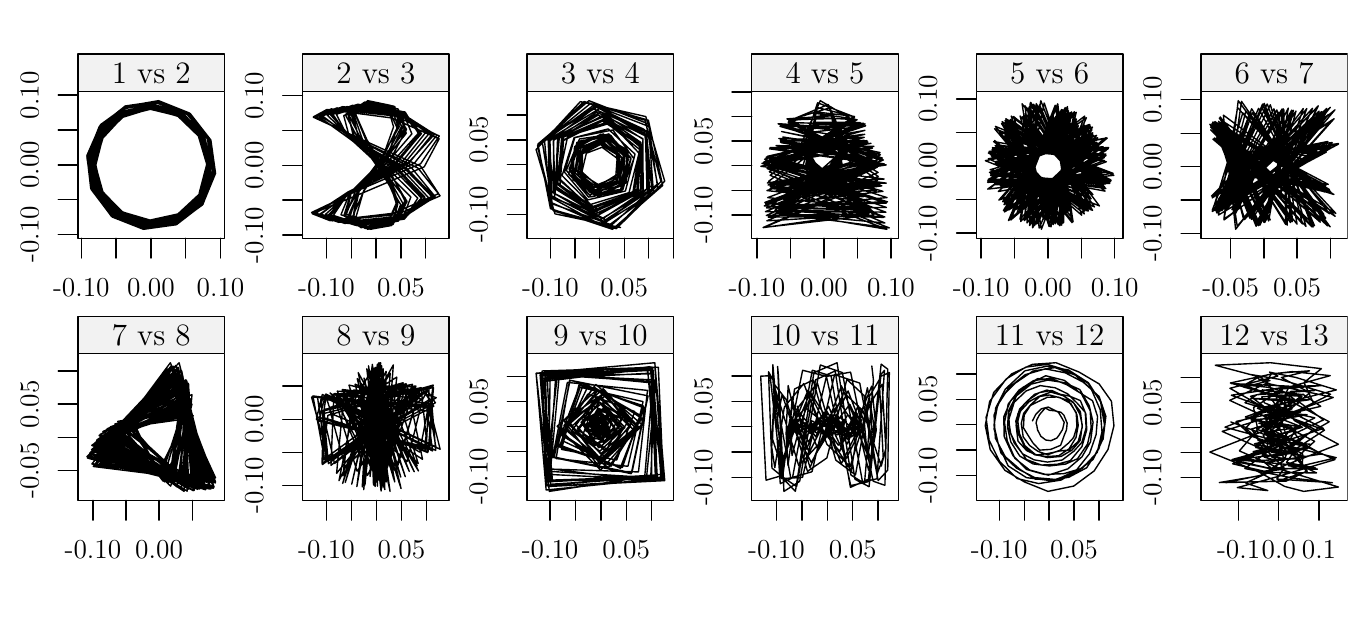}
    \caption{``MotorVehicle'', 2nd stage: scatterplots for eigenvector pairs ($L=264$)}
    \label{fig:peigvec}
\end{figure}

\begin{figure}[!htb]
    \centering
    \includegraphics{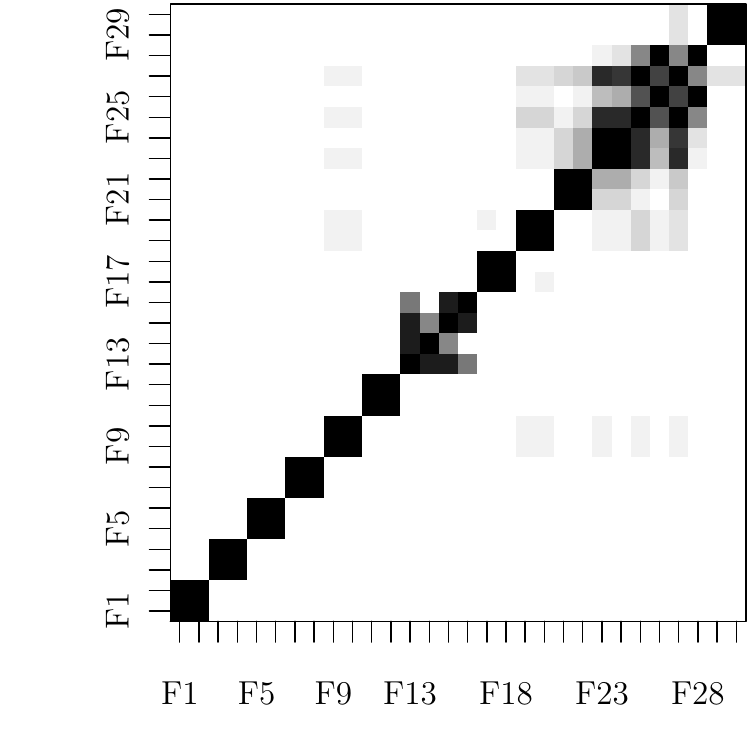}
    \caption{``MotorVehicle'', 2nd stage: $\bfw$-correlation matrix ($L=264$)}
    \label{fig:wcor}
\end{figure}

For proper identification of the sought sine waves, we will use the graph of
eigenvalues, scatterplots of eigenvectors and $\bfw$-correlation matrix of the
elementary components.  In Fig.~\ref{fig:eigvseason} we see several steps
produced by approximately equal eigenvalues. Each step is likely to be yielded
by a pair of eigenvectors which correspond to a sine
wave. Fig.~\ref{fig:peigvec} confirms our guess. One can see six almost regular
polygons.  ET1--2, ET3--4, ET5--6, ET7--8 and ET9--10 correspond to periods 12,
6, 2.4, 3, 4, which are produced by the seasonality and are clearly explained by
the periodogram (Fig.~\ref{fig:restrpgram}).  The components are ordered in
accordance with the ordering of the periodogram values at these frequencies.
Fig~\ref{fig:wcor} shows that the considered pairs of components are highly
correlated within and are almost not correlated between.  Note that there is one
more pair of eigentriples ET11--12 which satisfies the same properties.  Since
this pair corresponds to the period 16, which is not interpretable for monthly
data, we refer it to noise. The estimation of periods was performed by the
function \code{parestimate} and the results are
\begin{verbatim}
> parestimate(s, 1:12, method = "esprit-ls")$periods
 [1]  2.996167  -2.996167  12.008794 -12.008794  2.398670  -2.398670
 [7] 16.198097 -16.198097   5.982904  -5.982904  4.014053  -4.014053
> parestimate(s, 11:12, method = "pairs")
 [1] 15.9677
\end{verbatim}

Let us present the results of the series decomposition.

\begin{fragment}
\caption{``MotorVehicle'', 2nd stage: reconstruction and plotting of the results}
\begin{verbatim}
res2 <- reconstruct(s2, groups=list(1:10))
seasonality <- res2$F1;
res <- residuals(res2);
# Extracted seasonality
plot(res2, add.residuals = FALSE, col = c("black", "red"))
# Result of Sequential SSA
plot(res2, base.series = res1)
# Seasonally adjusted series
plot(MotorVehicle-seasonality, type='l')
\end{verbatim}
\end{fragment}

\begin{figure}[!htb]
    \centering
    \includegraphics{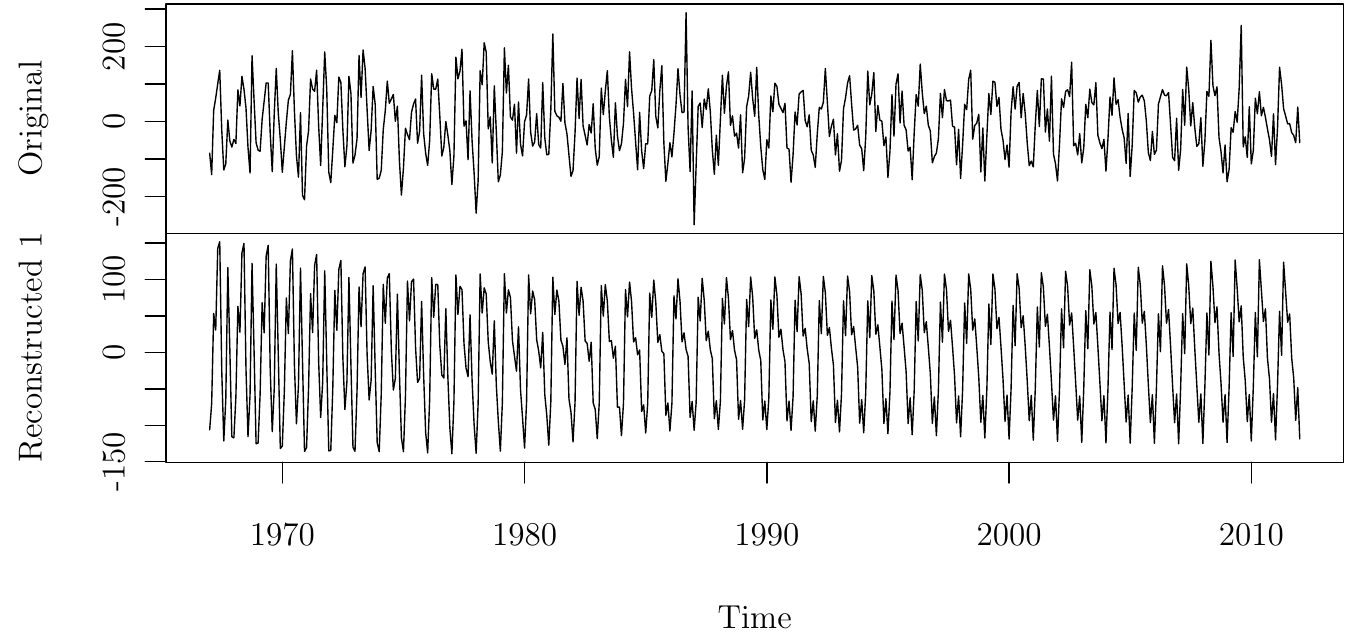}
    \caption{``MotorVehicle'', 2nd stage: the series and the extracted seasonal component}
    \label{fig:serper}
\end{figure}

\begin{figure}[!htb]
    \centering
    \includegraphics{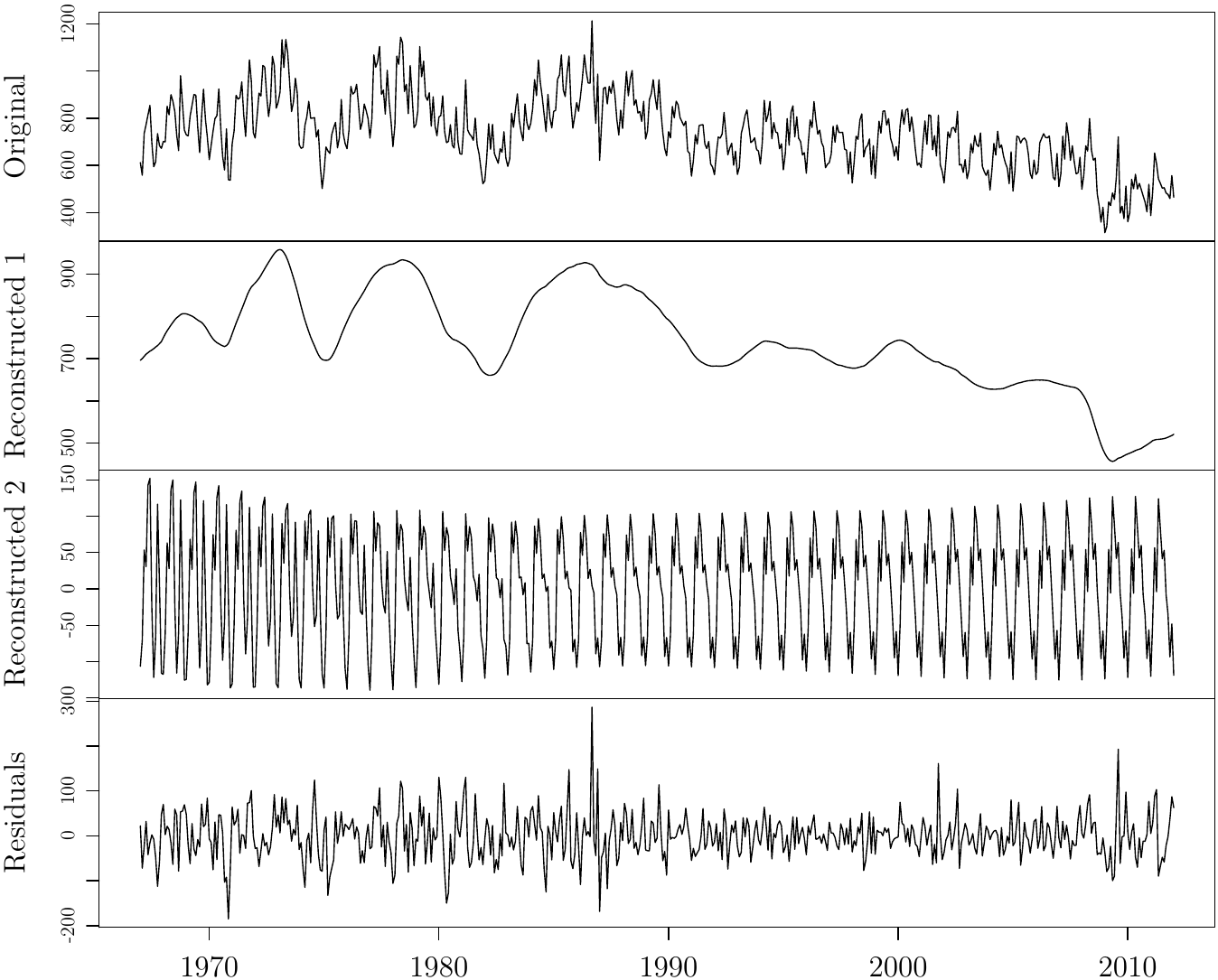}
    \caption{``MotorVehicle'': Series and its trend-periodic-residuals decomposition}
    \label{fig:allinone}
\end{figure}

\begin{figure}[!htb]
    \centering
    \includegraphics{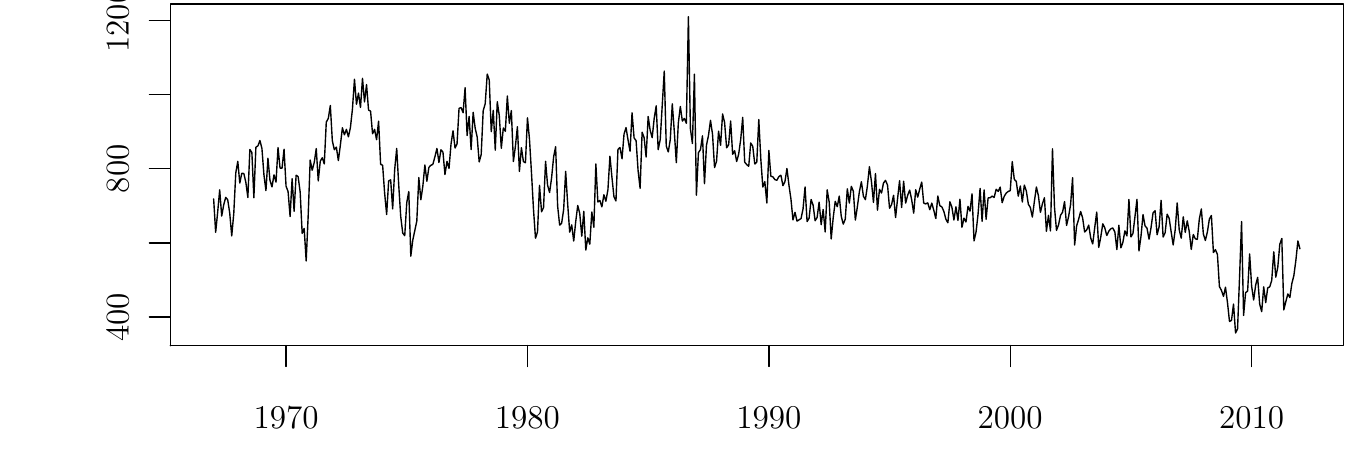}
    \caption{``MotorVehicle'': Seasonally adjusted series}
    \label{fig:seas_adjusted}
\end{figure}

The extracted seasonality (ET1--10) is depicted in Fig~\ref{fig:serper}. Slow
change of sine wave phases seen in Fig.~\ref{fig:peigvec} yields a periodic
behavior of complex form.  Fig.~\ref{fig:allinone} shows the resultant
decomposition of both stages of Sequential SSA.  Note that the obtained noise
residuals are heterogeneous.  As an auxiliary result, we obtain also seasonally
adjusted series (Fig.~\ref{fig:seas_adjusted}).

\begin{figure}[htb]
    \centering
    \includegraphics{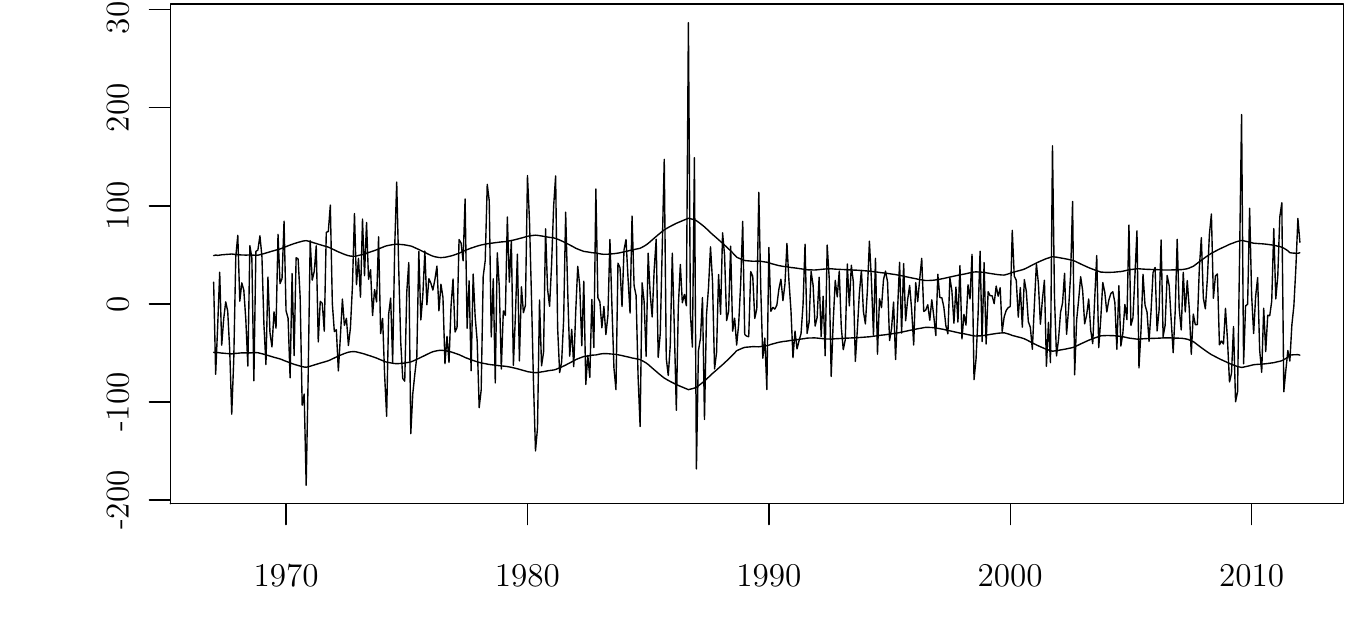}
    \caption{``MotorVehicle'': Residuals with envelopes}
    \label{fig:residuals}
\end{figure}

Finally, let us demonstrate how to estimate the variance of the heterogeneous
noise.  The procedure is based on two observations: first, the variance is equal
to the expectation of squared residuals; second, for stochastic process the
trend is its expectation.  Therefore, the variance can be estimated as the trend
of squared residuals.  This trend can be extracted by SSA with small window
length and reconstructed by the leading eigentriple. The choice of window length
makes affects the level of detail with which we see the extracted trend.  The
choice $L=30$ provides an appropriate trend. The result of
Fragment~\ref{an:noise} is depicted in Fig.~\ref{fig:residuals} containing the
residuals with standard deviation bounds.

\begin{fragment}
\caption{``MotorVehicle'': finding noise envelope}
\begin{verbatim}
s.env <- ssa(res^2, L=30)
rsd <- sqrt(reconstruct(s.env, groups=list(1))$F1)
plot(res, type='l'); lines(rsd, type='l'); lines(-rsd, type='l')
\end{verbatim}
\label{an:noise}
\end{fragment}

\begin{remark}
    For stationary time series the use of Toeplitz SSA is appropriate. In the described
    example, it makes  no sense to apply Toeplitz SSA for trend extraction.
    Generally, it can be applied to extraction of seasonality
    via the call \code{s <- ssa(res.trend, L=264, kind="toeplitz-ssa")}.
    However, the result of decomposition is worse, since the seasonal behavior
    is changing in time.  Note that the ordering of eigentriples by eigenvalues
    of the matrix $\bfS$ and their contribution to the decomposition
    differ. The values of \code{s\$lambda} are equal to the contribution values,
    while the ordering is performed by eigenvalues. Therefore the graph
    \code{plot(s)} can be not-monotonic.
\end{remark}

\section{SSA forecasting with R}
\label{sec:for}

\subsection{Typical code}
After the decomposition has been performed forecasting becomes
available. \pkg{Rssa} implements two methods of forecasting, recurrent and
vector forecasts.

\begin{fragment}
\caption{``co2'': forecasting}
\begin{verbatim}
# Decomposition stage
s <- ssa(co2, L = 120)
# Recurrent forecast, the result is the forecasted values only
# The result is the set of forecasts for each group
for1 <- rforecast(s, groups = list(1, c(1,4), 1:4, 1:6), len = 12)
matplot(data.frame(for1), type='b', pch = c('1','2','3','4'))
# Recurrent forecast, the forecasted points added to the base series
for1a <- rforecast(s, groups = list(1, c(1,4), 1:4, 1:6), len = 36,
                  only.new = FALSE)
# Plot of the forecast based on the second group c(1,4)
plot(cbind(co2, for1a$F2), plot.type='single', col=c('black','red'))
# Vector forecast
for2 <- vforecast(s, groups = list(1:6), len = 12, only.new = FALSE)
plot(cbind(co2, for2$F1), plot.type='single', col=c('black','red'))
# Confidence intervals, they can be calculated for one group only
for3 <- bforecast(s, group = 1:6, len = 12, type = "recurrent")
plot(for3, plot.type="single", col=c("black","red","red"))
\end{verbatim}
\label{fr:co2forecast}
\end{fragment}

Alternatively, one can use the all-in-one function \code{forecast} which serves
as a wrapper over \code{rforecast}, \code{vforecast} and \code{bforecast} and
yields the output compatible with the \pkg{forecast} package
\citep{Hyndman2012}. This way, one can use convenient graphic tools implemented
in the \pkg{forecast} package to plot the forecast results.  See
Fragment~\ref{for:2bootstrap} as an example of this.

Like the \code{reconstruct} function, all the forecasting routines try to use
the attributes of the initial series for the resulting series (in particular,
they try to add to the result the time index of the series). Unfortunately, this
cannot be done in class-neutral way as it is done in the \code{reconstruct} case
and needs to be handled separately for each possible type of time series. The
forecasting routines know how to impute the time indices for some standard time
series classes like \code{ts} and \code{zooreg}.

\begin{figure}[htb]
    \centering
    \includegraphics{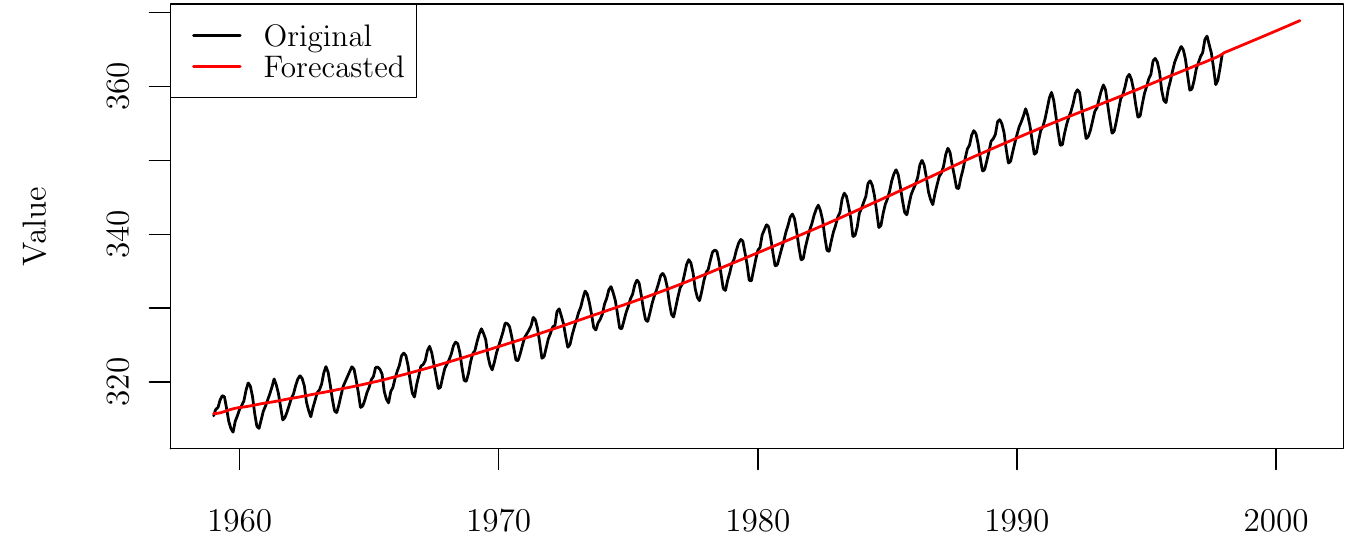}
    \caption{``co2'': trend forecast}
    \label{fig:co2forecast}
\end{figure}

The forecast for trend (ET1 and ET4) is shown on Fig.~\ref{fig:co2forecast}
together with the initial series.

In addition to forecasting, this block of \pkg{Rssa} functions provides tools
for analyzing the forecasting LRR (Fragment~\ref{fr:lrr}). The roots are ordered
by modulus since typically (but not always) the signal roots have maximal moduli
(see \citet{Usevich2010} for theoretical results about signal and extra roots).

\begin{fragment}
\caption{``co2'': linear recurrence relation}
\begin{verbatim}
num <- 1:6
lrr.coef<-lrr(s, group = num)
r <- roots(lrr.coef)
# Plot of roots against the unit circle
plot(lrr.coef)
\end{verbatim}
\label{fr:lrr}
\end{fragment}

For description of the forecast behavior, the signal roots of the
characteristic polynomial of the forecasting LRR and their parametric form
should be found.  For parameter estimation (frequency and damping rate) code of
Fragment~\ref{fr:parest} can be used.

\begin{fragment}
\caption{``co2'': parameter estimation}
\begin{verbatim}
print(2*pi/Arg(r[num]))
print(Mod(r[num]))
parestimate(s, 1:6, method = "esprit-ls")
parestimate(s, c(2:3,5:6), method = "esprit-ls")
\end{verbatim}
\label{fr:parest}
\end{fragment}

The result of estimation through the LRR roots is
\begin{verbatim}
> print(2*pi/Arg(r[num]))
[1] 5.999366  -5.999366  11.996071 -11.996071      Inf      Inf
> print(Mod(r[num]))
[1] 1.000575   1.000575   1.000385   1.000385 1.000354 0.985554
\end{verbatim}
All these roots are likely to be related to the signal. Results of application
of the ESPRIT confirm this.  This means that the explicit form of the forecast
is the sum of half of year and annual sine waves with almost constant amplitude
and also a trend approximated by the sum of two exponentials.

To find proper parameters of the method, testing of the forecasting formulas can
be performed.  The code from Fragment~\ref{for:sliding} shows how to implement
the function to see the dependence of the forecasting accuracy on the window
length and on the number of the selected components.

\begin{fragment}
\caption{Function for sliding forecasts}
\begin{verbatim}
forecast.check <- function(F,
                           groups,
                           forecast.len = 1, sliding.len = N %/% 4,
                           ...,
                           type = c("recurrent", "vector")) {
  type <- match.arg(type)

  N <- length(F)
  K.sliding <- N - sliding.len - forecast.len + 1

  r <- matrix(nrow = K.sliding, ncol = length(groups))
  f.fun <- if (identical(type, "vector")) vforecast else rforecast
  for (i in 1:K.sliding) {
    F.train <- F[seq(from = i, len = sliding.len)]
    F.check <- F[seq(from = sliding.len + i, len = forecast.len)]
    s <- ssa(F.train, ...)
    for (idx in seq_along(groups)) {
      group <- groups[[idx]]
      f <- f.fun(s, groups = list(group), len = forecast.len)[[1]]
      r[i, idx] <- mean((f - F.check)^2)
    }
  }
  apply(r, 2, mean)
}
\end{verbatim}
\label{for:sliding}
\end{fragment}

Fragment~\ref{for:accuracy} contains examples showing how to use sliding
forecasts for choosing the parameters.  Commented lines show other reasonable
choice for the corresponding variables.  Length of sliding subseries equals 360,
while the series length is equal to 468.  Choice \code{fl <- 1} corresponds to
108 one-step ahead forecasts (short-term forecasting), choice \code{fl <- 108}
corresponds to one 108-step ahead forecast (long-term forecasting).

\begin{fragment}
\caption{``co2'': dependence of forecast accuracy on choice of parameters}
\begin{verbatim}
Lmin <- 24; N <- length(co2); ns <- 360
fl <- N-ns # fl <- 1
# groups <- list(1:6, 1:10, 1:15, 1:20)
groups <- list(c(1,4), 1:4, 1:6, c(1:6, 14, 15))
Lseq <- seq(Lmin, ns-Lmin, by = 6)
fcL <- function(L) forecast.check(co2, groups,
                                  forecast.len=1, sliding.len = ns,
                                  L = L, neig = 20, type="vector")
m <- sapply(Lseq, fcL)
matplot(time(co2)[Lseq], t(m),
        type = "l", col=c("red","green","blue","black"))
\end{verbatim}
\label{for:accuracy}
\end{fragment}

\subsection{Case study}
Let us consider the same example ``MotorVehicle''.  Since the trend has complex
structure, it makes sense to forecast the trend and seasonality separately.

We start with forecasting the seasonality. Fragment~\ref{for:2for} performs
forecasts by recurrent and vector methods.
\begin{fragment}
\caption{``MotorVehicle'' seasonality: forecasting}
\begin{verbatim}
s1 <- ssa(MotorVehicle, L=12)
trend <- reconstruct(s1, groups = list(1))
res.trend <- residuals(trend); trend <- trend$F1
s2 <- ssa(res.trend, L=264)
frec <- rforecast(s2, groups = list(1:10), len = 60)$F1
fvec <- vforecast(s2, groups = list(1:10), len = 60)$F1
plot(cbind(frec, fvec), plot.type = "single", col=c("black","red"))
\end{verbatim}
\label{for:2for}
\end{fragment}

The results are stable enough and the difference between recurrent and vector
forecasts is very small. To estimate the forecasting error, the bootstrap
confidence intervals for the forecasted component can be calculated. This can be
done with the help of the function \code{bforecast} or of its wrapper
\code{forecast} for using the plotting facilities from \pkg{forecast} package,
see Fragment~\ref{for:2bootstrap}.

\begin{fragment}
\caption{``MotorVehicle'' seasonality: bootstrap confidence intervals}
\begin{verbatim}
f <- forecast(s2, group = 1:10, len = 60,
              method = "bootstrap-recurrent")
plot(f, include = 60, shadecols = "green")
\end{verbatim}
\label{for:2bootstrap}
\end{fragment}

\begin{figure}[htb]
    \centering
    \includegraphics{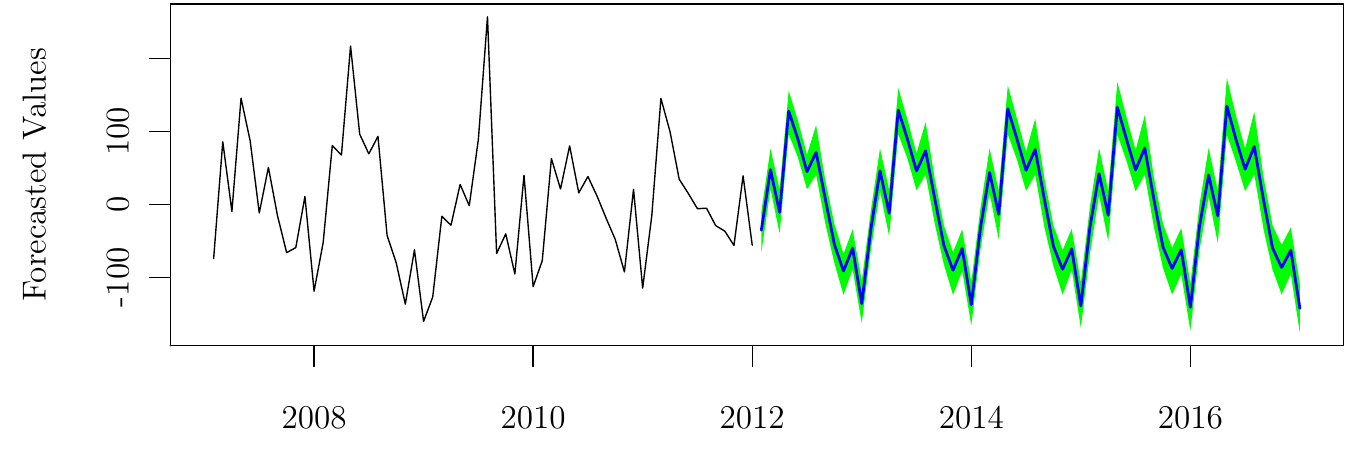}
    \caption{``MotorVehicle'' seasonality: bootstrap confidence intervals}
    \label{fig:2bootstrap}
\end{figure}

The confidence bounds are depicted in Fig.~\ref{fig:2bootstrap}. Certainly, at
least approximate independence and identical distribution of residuals should be
checked before using confidence intervals based on these assumptions.

\begin{figure}[htb]
    \centering
    \includegraphics{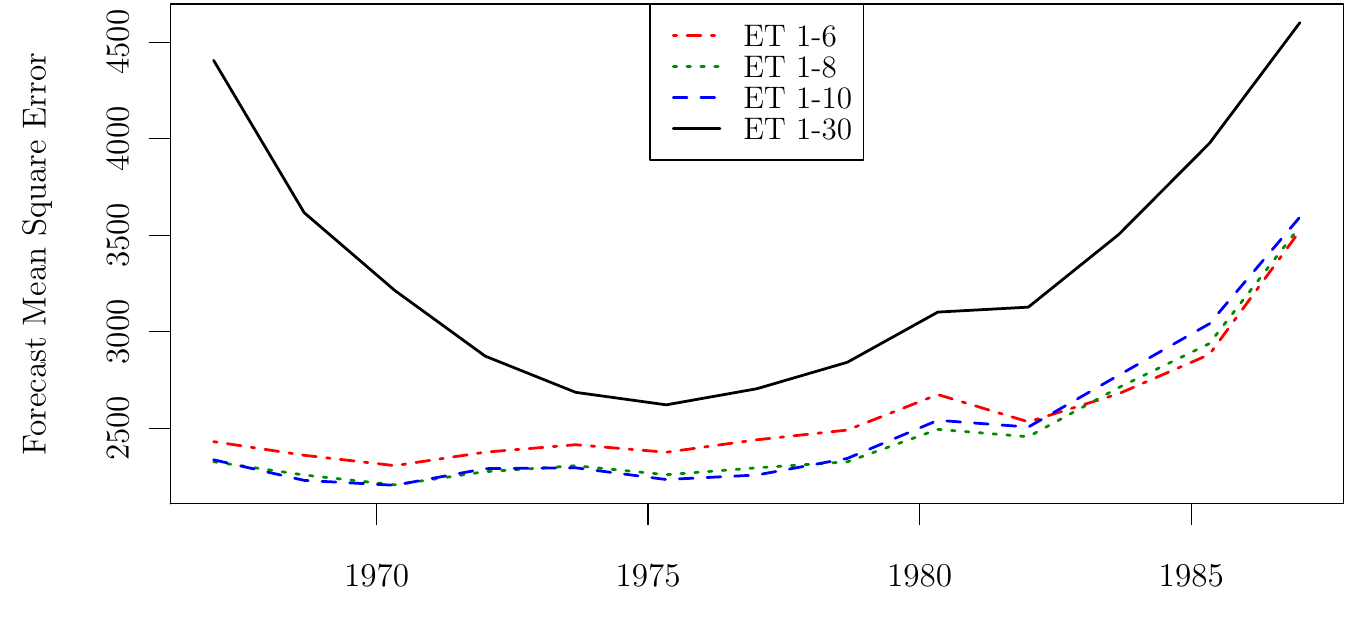}
    \caption{``MotorVehicle'' seasonality: dependence of forecast accuracy on number of removed points}
    \label{fig:2sliding}
\end{figure}

Let us check, if the removal of the starting period of the time series can improve
accuracy.
\begin{fragment}
\caption{``MotorVehicle'' seasonality: dependence of forecast accuracy on number of removed points}
\begin{verbatim}
N <- length(MotorVehicle)
groups <- list(1:6, 1:8, 1:10, 1:30)
Nstart <- seq (1, 241, 20);
fcT <- function(NN)
  forecast.check(res.trend[NN:N], groups,
                 forecast.len = 12, sliding.len = 240,
                 L = 120, type="recurrent", svd.method="eigen")
m <- sapply(Nstart, fcT)
matplot(time(MotorVehicle)[Nstart], t(m),
        type = "l", col=c("red","green","blue","black"))
\end{verbatim}
\label{for:2sliding}
\end{fragment}
Fragment~\ref{for:2sliding} uses the function defined in
Fragment~\ref{for:sliding}.  One can see in Fig.~\ref{fig:2sliding} that it is
better to use the whole time series and perform the forecasting based on ET1--8
or ET1--10.

Totally different situation takes place with the trend forecast.  Since the
trend possibly has a structure changing in time, it is unreasonable to use the
whole trend for forecasting. Therefore, we need to find the last point of the
structure change.

Probably, the last point of structure change is 2009 (crisis).\footnote{The
    points of the structural change need to be studied additionally either via
    SSA heterogeneity matrix or using other methods of change-point analysis.}
Therefore, we make a forecast using the data from the last 3 years, that is, the
last 36 points (Fragment~\ref{for:1for1}).

\begin{fragment}
\caption{``MotorVehicle'' trend: forecasting of last 3 years behavior}
\begin{verbatim}
trend.end1 <- ts(trend[506:541],
                end = end(trend), frequency = frequency(trend))
s.end1 <- ssa(trend.end1)
frec1 <- rforecast(s.end1, groups = list(1),
                  len = 24, only.new = FALSE)$F1
plot(cbind(trend.end1, frec1), plot.type="single",
     col=c("black","red"))
\end{verbatim}
\label{for:1for1}
\end{fragment}

\begin{figure}[htb]
    \centering
    \includegraphics{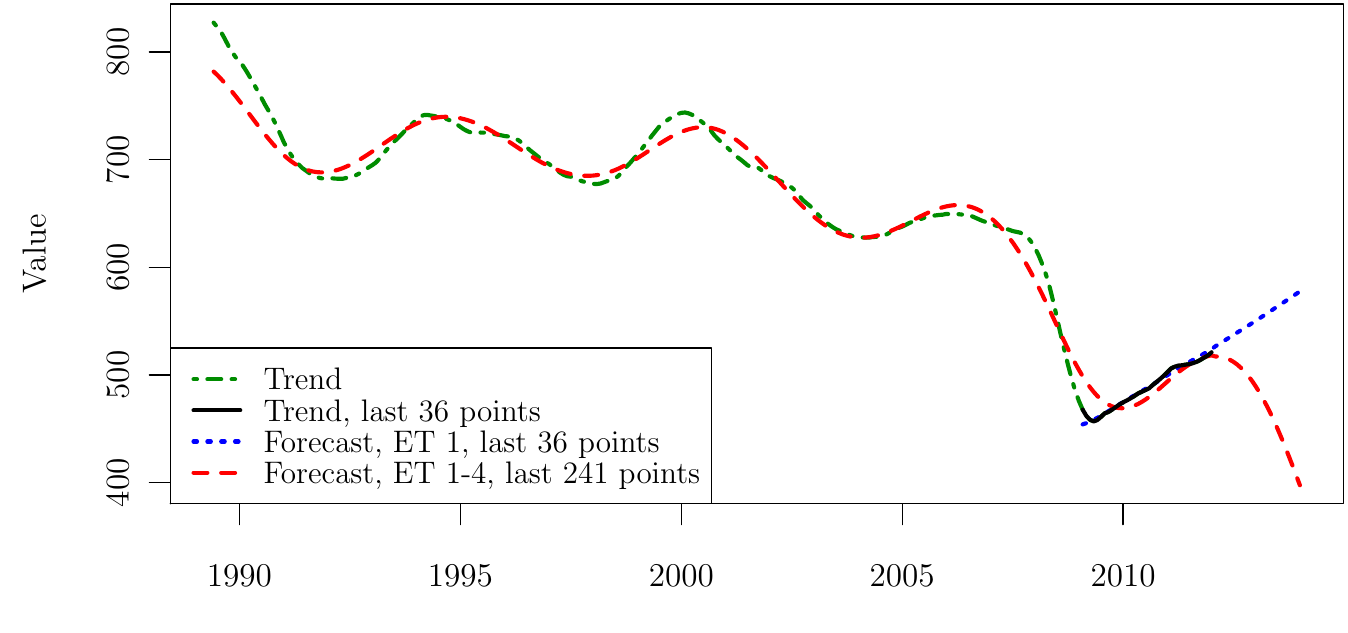}
    \caption{``MotorVehicle'' trend: forecasting of last 3 and 22 years behavior}
    \label{fig:1for2}
\end{figure}

If we consider a longer time period for forecasting (Fragment~\ref{for:1for2}),
then we will see a totally different forecast (Fig.~\ref{fig:1for2}).

\begin{fragment}
\caption{``MotorVehicle'' trend: forecasting of last 22 years behavior}
\begin{verbatim}
trend.end2 <- ts(trend[270:541],
                end = end(trend), frequency = frequency(trend))
s.end2 <- ssa(trend.end2)
frec2 <- rforecast(s.end2, groups = list(1:4),
                  len = 24, only.new = FALSE)$F1
plot(cbind(trend.end2, frec2), plot.type="single",
     col=c("black","red"))
\end{verbatim}
\label{for:1for2}
\end{fragment}

To choose the proper forecast, additional macroeconomic analysis is necessary.

\section*{Acknowledgement}
We thank the editor, both reviewers and Anatoly Zhigljavsky (Cardiff) for useful
comments, which helped to make the paper much clearer and the \R{} code more
standard and more user-friendly.  We are also grateful to Alex Shlemov
(St.Petersburg) for his participation in the \pkg{Rssa} development, especially
for his fast implementation of the vector forecasting algorithm.

\bibliographystyle{elsarticle-harv}
\bibliography{rssa}

\begin{thebibliography}{20}
\expandafter\ifx\csname natexlab\endcsname\relax\def\natexlab#1{#1}\fi
\expandafter\ifx\csname url\endcsname\relax
  \def\url#1{\texttt{#1}}\fi
\expandafter\ifx\csname urlprefix\endcsname\relax\def\urlprefix{URL }\fi

\bibitem[{Anderson et~al.(1999)Anderson, Bai, Bischof, Blackford, Demmel,
  Dongarra, Du~Croz, Greenbaum, Hammarling, McKenney, and Sorensen}]{Lapack99}
Anderson, E., Bai, Z., Bischof, C., Blackford, S., Demmel, J., Dongarra, J.,
  Du~Croz, J., Greenbaum, A., Hammarling, S., McKenney, A., Sorensen, D., 1999.
  {LAPACK} Users' Guide, 3rd Edition. SIAM, Philadelphia, PA.

\bibitem[{{Badeau} et~al.(2005){Badeau}, {Richard}, and {David}}]{Badeau05}
{Badeau}, R., {Richard}, G., {David}, B., 2005. Fast adaptive {ESPRIT}
  algorithm. In: Statistical Signal Processing'05. Bordeaux, France, pp.
  289--294.

\bibitem[{Elsner and Tsonis(1996)}]{Elsner.Tsonis1996}
Elsner, J.~B., Tsonis, A.~A., 1996. {S}ingular {S}pectrum {A}nalysis: A New
  Tool in Time Series Analysis. Plenum.

\bibitem[{Ghil et~al.(2002)Ghil, Allen, Dettinger, Ide, Kondrashov, Mann,
  Robertson, Saunders, Tian, Varadi, and Yiou}]{Ghil.etal2002}
Ghil, M., Allen, R.~M., Dettinger, M.~D., Ide, K., Kondrashov, D., Mann, M.~E.,
  Robertson, A., Saunders, A., Tian, Y., Varadi, F., Yiou, P., 2002. Advanced
  spectral methods for climatic time series. Rev. Geophys. 40~(1), 1--41.

\bibitem[{Golub and Van~Loan(1996)}]{Golub.VanLoan1996}
Golub, G.~H., Van~Loan, C.~F., 1996. {Matrix computations (3rd ed.)}. Johns
  Hopkins University Press, Baltimore, MD, USA.

\bibitem[{Golyandina(2010)}]{Golyandina2010}
Golyandina, N., 2010. On the choice of parameters in singular spectrum analysis
  and related subspace-based methods. Stat. Interface 3~(3), 259--279.

\bibitem[{Golyandina et~al.(2001)Golyandina, Nekrutkin, and
  Zhigljavsky}]{Golyandina.etal2001}
Golyandina, N., Nekrutkin, V., Zhigljavsky, A., 2001. Analysis of Time Series
  Structure: {SSA} and Related Techniques. Chapman\&Hall/CRC.

\bibitem[{Golyandina et~al.(2012)Golyandina, Pepelyshev, and
  Steland}]{Golyandina.etal2012}
Golyandina, N., Pepelyshev, A., Steland, A., 2012. New approaches to
  nonparametric density estimation and selection of smoothing parameters.
  Comput. Stat. Data Anal. 56~(7), 2206--2218.

\bibitem[{Golyandina and Zhigljavsky(2013)}]{Golyandina.Zhigljavsky2012}
Golyandina, N., Zhigljavsky, A., 2013. {S}ingular {S}pectrum {A}nalysis for
  time series. Springer Briefs in Statistics. Springer.

\bibitem[{Hassani et~al.(2009)Hassani, Heravi, and
  Zhigljavsky}]{Hassani.etal2009}
Hassani, H., Heravi, S., Zhigljavsky, A., 2009. Forecasting {E}uropean
  industrial production with singular spectrum analysis. Int. J. Forecast.
  25~(1), 103 -- 118.

\bibitem[{Hyndman(2012)}]{Hyndman2012}
Hyndman, R.~J., 2012. forecast: Forecasting functions for time series and
  linear models. R package version 3.25, with contributions from Slava Razbash
  and Drew Schmidt.

\bibitem[{Jenkins and Traub(1970)}]{Jenkins1970}
Jenkins, M., Traub, J., 1970. A three-stage variable-shift iteration for
  polynomial zeros and its relation to generalized rayleigh iteration.
  Numerische Mathematik 14, 252--263.

\bibitem[{Keeling and Whorf(1997)}]{Keeling1997}
Keeling, C.~D., Whorf, T.~P., 1997. Atmospheric {CO2} concentrations ---
  {M}auna {L}oa {O}bservatory, {H}awaii, 1959-1997. Scripps Institution of
  Oceanography (SIO), University of California, La Jolla, California USA
  92093-0220.

\bibitem[{Korobeynikov(2010)}]{Korobeynikov2010}
Korobeynikov, A., 2010. Computation- and space-efficient implementation of
  {SSA}. Stat. Interface 3~(3), 357--368.

\bibitem[{Larsen(1998)}]{Larsen98}
Larsen, R.~M., 1998. Efficient algorithms for helioseismic inversion. Ph.D.
  thesis, University of Aarhus, Denmark.

\bibitem[{{Roy} and {Kailath}(1989)}]{Roy.Kailath1989}
{Roy}, R., {Kailath}, T., 1989. {ESPRIT: estimation of signal parameters via
  rotational invariance techniques}. IEEE Trans. Acoust. 37, 984--995.

\bibitem[{{U.S. Bureau of Economic Analysis}(2012)}]{MotorVehicle2012}
{U.S. Bureau of Economic Analysis}, 2012. Table 7.2.5s. {A}uto and {T}ruck
  {U}nit {S}ales {P}roduction {I}nventories {E}xpenditures and {P}rice.

\bibitem[{Usevich(2010)}]{Usevich2010}
Usevich, K., 2010. On signal and extraneous roots in {Singular Spectrum
  Analysis}. Stat. Interface 3~(3), 281--295.

\bibitem[{Yamazaki et~al.(2008)Yamazaki, Bai, Simon, Wang, and Wu}]{Yamazaki08}
Yamazaki, I., Bai, Z., Simon, H., Wang, L.-W., Wu, K., 2008. Adaptive
  projection subspace dimension for the thick-restart {L}anczos method. Tech.
  rep., Lawrence Berkeley National Laboratory, University of California, One
  Cyclotron road, Berkeley, California 94720.

\bibitem[{Zeileis and Grothendieck(2005)}]{Zeileis.Grothendieck2005}
Zeileis, A., Grothendieck, G., 2005. zoo: {S3} infrastructure for regular and
  irregular time series. Journal of Statistical Software 14~(6), 1--27.

\end{thebibliography}

\end{document}